\documentclass[twoside,final,onecolumn]{jfm}

\usepackage{newtxtext}
\usepackage{newtxmath}
\usepackage{natbib}
\usepackage{amssymb}
\usepackage{hyperref}
\hypersetup{
    colorlinks = true,
    urlcolor   = blue,
    citecolor  = blue, %Black
}

\newcommand{\RomanNumeralCaps}[1]

\usepackage{xcolor}
\usepackage{amsmath}
\usepackage{rotating}
\usepackage{graphicx}
\usepackage{soul}
\usepackage{color}
\definecolor{mycolor}{RGB}{ 186, 219, 237}
   
\usepackage{upgreek}
\newcommand{\RNum}[1]{\uppercase\expandafter{\romannumeral #1\relax}}
\shorttitle{Localized performance of riblets in boundary layers past finite length bodies}
\shortauthor{S. Fu, and S. Raayai-Ardakani}

\title{Localized performance of riblets with curved cross-sectional profiles in boundary layers past finite length bodies} 

\author{Shuangjiu Fu\aff{1}
 \and Shabnam Raayai-Ardakani\aff{1} \corresp{\email{sraayai@fas.harvard.edu}}}

\affiliation{\aff{1}Rowland Institute, Harvard University,
60 Oxford St., Cambridge, MA 02138, USA}

\begin{document}

\maketitle

\begin{abstract}
Riblets are a well-known passive drag reduction technique with the potential for as much as 9\% reduction in the frictional drag force in laboratory settings, and proven benefits for large scale aircraft. However, less information is available on the applicability of these textures for smaller air/waterborne vehicles where assumptions such as periodicity and/or asymptotic nature of the boundary layer no longer apply and the shape of the bodies of these vehicles can give rise to moderate levels of pressure drag. Here, we explore the effect of riblets on both sides of a finite-size foil consisting of a streamlined leading edge and a flat body. We use high resolution two-dimensional, two-component particle image velocimetry, with a double illumination and consecutive-overlapping imaging technique to capture the velocity field in both the boundary layer and the far field. We find the local velocity profiles and shear stress distribution, as well as the frictional and pressure components of the drag force and show the possibility of achieving reduction in both the fictional and pressure components of the drag force and record cumulative drag reduction as much as $6\%$. We present the intertwined relationship between the distribution of the spanwise-averaged shear stress distribution, the characteristics of the velocity profiles, and the pressure distribution around the body, and how the local distribution of these parameters work together or against each other in enhancing or diminishing the drag-reducing ability of the riblets for the entirety of the body of interest. 
\end{abstract}

\begin{keywords}
Boundary layer control, Drag reduction
% Authors should not enter keywords on the manuscript, as these must be chosen by the author during the online submission process and will then be added during the typesetting process (see http://journals.cambridge.org/data/\linebreak[3]relatedlink/jfm-\linebreak[3]keywords.pdf for the full list)
\end{keywords}

\section{Introduction} \label{intro}

Riblets, consisting of streamwise periodic grooves, are a well-known passive technique to reduce the drag on surfaces where experiments and simulations have reported maximums of between 6-9\% reductions in the skin friction drag in a variety of conditions \citep{walsh1983riblets, walsh1984optimization, bechert1997experiments, raayai2017drag, raayai2019geometric, raayai2020geometry, raayai2022polynomial, wong2024viscous, garcia2011drag, viswanath1995turbulent, dinkelacker1988possibility, chamorro2013drag, bechert2000fluid, wong2024viscous}. In larger scale experiments with model or full-scale vehicles, riblets have also proven to be effective; In tests with a High Speed Buoyancy Propelled Vehicle (MOBY-D), \citet{choi1990drag} reported a 3.4\% reduction in the frictional drag, and \citet{szodruch1991viscous} showed that an Airbus 320, covered with 70\% riblets can deliver almost a 2 \% reduction in total drag.  \citet{walsh1989riblet} also report that %in a separate test, 
a V-groove plate of 5.83 ft by 1 ft placed at about 6.2 ft aft of the nose of a modified Gates Learjet model 28/29 twin-jet business jet experienced a maximum of 6\% reduction as measured by on-board boundary layer (BL) wakes attached to the plane. 

Two mechanisms have been identified for the drag reduction with riblets; The first one, applicable to both laminar and turbulent flows, is the presence of slow and near quiescent flow inside the grooves \citep{bacher1986turbulent, raayai2017drag, raayai2019geometric, wallace1988viscous, chu1993direct, djenidi1994laminar, baron1993boundary}. This slow moving fluid causes the shear stress inside the grooves to be lower than that of the smooth reference and only close to the peaks of the grooves, shear stresses larger than the smooth surface is reported \citep{khan1986numerical, park1994flow, raayai2019geometric, raayai2022polynomial, modesti2021dispersive, choi1993direct, chu1993direct}. This spanwise shear stress distribution can reduce the average shear stress experienced by the riblets and result in an overall reduction in the frictional drag. The second mechanism, only applicable to turbulent flows, is related to the ability of the drag-reducing riblets to adjust the spanwise flow and quasi-streamwise turbulent vortices \citep{djenidi1996laser, goldstein1995direct, choi1993direct, lee2001flow}. \citet{djenidi1996laser} report that drag-reducing riblets experience weaker quasi-streamwise vortices compared to the smooth wall which results in a less effective transport of momentum to the wall. \citet{goldstein1995direct} report that the quasi-streamwise vortices are pushed away from the wall in the presence of drag-reducing riblets and \citet{el2007drag} show that vortices penetrate rectangular drag-increasing riblets. \citet{viswanath2002aircraft} and \citet{caram1991effect} report lower level of Reynolds shear stress close to the wall and \citet{lancey1989effects} measured reductions in the velocity and pressure fluctuations at the wall. Taking the combined importance of the flow retardation inside the grooves and the relocation of the cross-sectional flow into account, \citet{luchini1991resistance} and \citet{luchini1995asymptotic} present a linear protrusion height model where the ability of the riblets in reducing/increasing the drag is related to the difference between the apparent origin of the streamwise flow inside the grooves and the origin of the spanwise flow,  both obtained using Stokes-flow calculations for a given riblet \citep{luchini1995asymptotic, bechert1997experiments, gruneberger2011drag}. \citet{wong2024viscous} have taken the modelling further and offer a viscous vortex model where the turbulent scale is allowed to interact with a non-vanishing riblet size to accurately predict the drag performance of the riblets in the viscous limit (using direct numerical simulations).  

Riblets as drag-reducing agents affect the near-wall BL and thus have been mainly explored as a means of reducing the skin friction drag, especially in zero pressure gradient conditions \citep{walsh1979drag, wong2024viscous, bechert1997experiments, endrikat2022reorganisation, choi1993direct, chu1993direct,abu_rowin_ghaemi_2019, dinkelacker1988possibility, chamorro2013drag, hou2017three}, with only reports of a few cases for well-defined adverse and favorable pressure gradient BL \citep{debisschop1996turbulent, klumpp2010riblets, nieuwstadt1993reduction, choi1990effects} and wake studies \citep{caram1989effects, caram1991effect, caram1992development} existing so far. The impact of riblets on the pressure distribution, pressure drag, and, if applicable, lift, is relatively unexplored \citep{van1988drag, nieuwstadt1993reduction, choi1990effects}. The ability of riblets in altering the pressure distribution around an airfoil has been recently shown using numerical simulations \citep{mele2012numerical, mele2020effect} where the changes in the shear stress of a riblet-covered aircraft  compared with a smooth body are a function of the position along the aircraft \citep{mele2016performance}. In addition, under the pressure distribution caused by a constant angle of attack of $2.25^{\circ}$, riblets offer a $4\%$ reduction in drag and a $5\%$ increase in lift. However, to keep the lift coefficient constant, the riblet-covered aircraft needs to be operated at a lower angle of attack of $2.09^{\circ}$, with potential for over $9\%$ reduction in drag \citep{mele2016performance}. 

In general, the levels of drag reduction captured by riblet surfaces are a function of both the dynamics of the flow (Reynolds number), and the geometry of the textures (spacing, height, and shape of the textures). The effect of the spacing of the riblets in turbulent wall units have been considered the longest in the literature since the initial work of \citet{walsh1979drag} and the effect of the height in the turbulent wall units have been added in the later work of \citet{walsh1983riblets}. However, the effect of the cross-sectional shape of the textures have been generally studied in a more qualitative manner with the tested shapes presented in forms of visual drawings in the reports \citep{wong2024viscous, bechert1997experiments, walsh1980drag, walsh1984optimization, rouhi2022riblet, modesti2021dispersive, choi1989tests}. Through the definition of the $\ell_g^+$, or the dimensionless cross-sectional area of the textures in wall units, as suggested by \citet{garcia2011drag}, the effect of shape has been considered in a more quantitative manner. This definition is successful in shifting the experimental and numerical results formerly presented in terms of either the spacing or height of the textures in the wall units along abscissa of the reduction curves and collapsing them into nearly similar curves with the maximum levels of reductions taking place at $\ell_g^+ \approx 10.7$. The remaining differences in the responses of different riblet profiles are only visible in the value of the reductions reported along the ordinate of the plots \citep{garcia2011drag, wong2024viscous, endrikat2022reorganisation, garcia2019control}. While the $\ell_g^+ \approx 10.7$ offers a physical parameter to identify the geometric guideline and operating conditions to get the best performance from the riblets, it does not offer any way to identify the best shape(s) to get the largest reduction for all possible riblet profiles with $\ell_g^+ \approx 10.7$. To characterize the shape of the cross-sectional profile of the textures in a unique quantitative manner, \citet{raayai2022polynomial} introduced a polynomial framework where the shape of the texture is defined using a vector of geometric parameters $\boldsymbol{\kappa} = $ [ ${\cal R} = -\kappa_1, \kappa_2, ..., \kappa_j$] using ${n_{\rm w}(z)}/{(\lambda/2)} = \sum_{j=0}^J \sum_{l=j}^J m_{jl}\kappa_l \left({z}/{(\lambda/2)}\right)^j$ where $m_{jl}$ are constant coefficients, $\lambda$ is the spacing of the riblets, and $\kappa_1 = -{\cal R}$ with ${\cal R} = A/(\lambda/2)$, the height-to-half-spacing ratio.

Here, we aim to investigate the possibility of the use of riblet surfaces on smaller vehicles, such as the case of smaller unmanned aerial/underwater vehicles that operate at high Reynolds number laminar conditions, where the assumption of the periodicity and self-similarity in the streamwise direction is not readily applicable. In laminar flows, the asymptotic protrusion height model of \citet{luchini1995asymptotic} for zero-pressure gradient self-similar Blasius BL and using Stokes flow formulation for the the flow inside the grooves, predicts a slight increase in the drag force. A domain perturbation and asymptotic expansion formulation of the Blasius equations solved using conformal mapping is able to predict up to about 1\% of reduction possible for textures with height-to-half-spacing of less than 0.8 and ${\rm Re}_L (\lambda/L)^2<1$ \citep{raayai2019geometric}. Numerical and experimental work of \citet{djenidi1989numerical} and \citet{djenidi1994laminar} and the numerical simulations of \citet{raayai2017drag} have shown the possibility of achieving larger drag reductions in laminar BL with riblets. 

The streamwise development of a BL along a fully textured flat plate and prior to transition to turbulence \citep{raayai2017drag, raayai2019geometric} (as opposed to that of the axisymmetric Taylor-Couette flows \citep{raayai2020geometry, raayai2022polynomial, greidanus2015turbulent, greidanus2015riblet}, or fully developed channel flows \citep{choi1993direct, goldstein1995direct} which are independent of the streamwise direction) results in the shear stress response of the surfaces to also depend on the local streamwise direction or the local Reynolds number. Theoretical and numerical studies of \citet{raayai2017drag, raayai2019geometric} using fully textured plates with constant cross-sectional profiles show that the spanwise-averaged shear stress, $\tau^*$ (normalized with the wetted cross-sectional contour length), starts at nearly the same value as that of the smooth reference plate at the leading edge of the plate and as the BL develops along the plate, $\tau^*$ decreases more than that of the smooth surface. However, the increase in the wetted surface area of the riblets (compared to the smooth) results in the riblets to be drag- (and shear-) increasing for certain plate lengths and only become drag-reducing past a point corresponding to ${\rm Re}_L (\lambda/L)^2<1$).

With this background, here we apply riblets of different cross-sectional shapes on both sides of a finite-sized symmetric hydrofoil operated at global Reynolds numbers of ${\rm Re}_L = 12,200$, $18,500$, and $24,200$ and evaluate both the local and global performance of the riblets in modulating the skin-friction coefficient and the components of the drag force. The results are presented in comparison against a smooth reference hydrofoil. More details of the reference case have been previously reported by \citet{fu2023doublelightsheet}. Here, we evaluate the performance of standalone riblet surfaces on the entirety of the body of the sample hydrofoils with finite thickness and curved leading edges, with a more applied approach and step away from the earlier experimental formats such as the use of one-sided samples \citep{endrikat2022reorganisation, walsh1983riblets, bacher1986turbulent}, partial riblet-coverage \citep{grek1996experimental}, and samples installed as part of the tunnel wall \citep{walsh1983riblets}, or localized high-resolution tomographic particle image velocimetry (PIV) in the middle of a flat riblet \citep{hou2017three}, as well as the setups of the numerical simulations employing periodic domains \citep{chu1993direct, goldstein1995direct, rouhi2022riblet, wong2024viscous, choi1993direct}, which have been instrumental in advancing the physical understanding of the effect of riblets on flow. 

This paper is thus organized as below: In section \ref{sec:methods}, we discuss the design of the riblets and samples, as well as the experimental procedure and data analysis. In section \ref{sec:results}, we first explore the global performance of the riblets in terms of the total drag, and then take advantage of a 3-tiered force measurement technique to decompose the total drag in terms of friction and pressure drag, as well as contribution from other factors not considered here. We discuss the differences as a function of the shape and sizes of the riblets. Finally, we investigate the local flow behaviour in the presence of the riblets and how the local shear stress and pressure distributions affect the global drag force experienced by the riblet covered samples. In section \ref{sec:conclusion}, we provide a summary of our finding and outline how this localized view of the flow can be used to enhance the designs of drag-reducing riblets for smaller vehicles.

\section{Methods} \label{sec:methods}

\subsection{Riblet Geometry} \label{sec:riblet_geom}

\begin{figure}
    \centering
    \includegraphics[width = 1\textwidth]{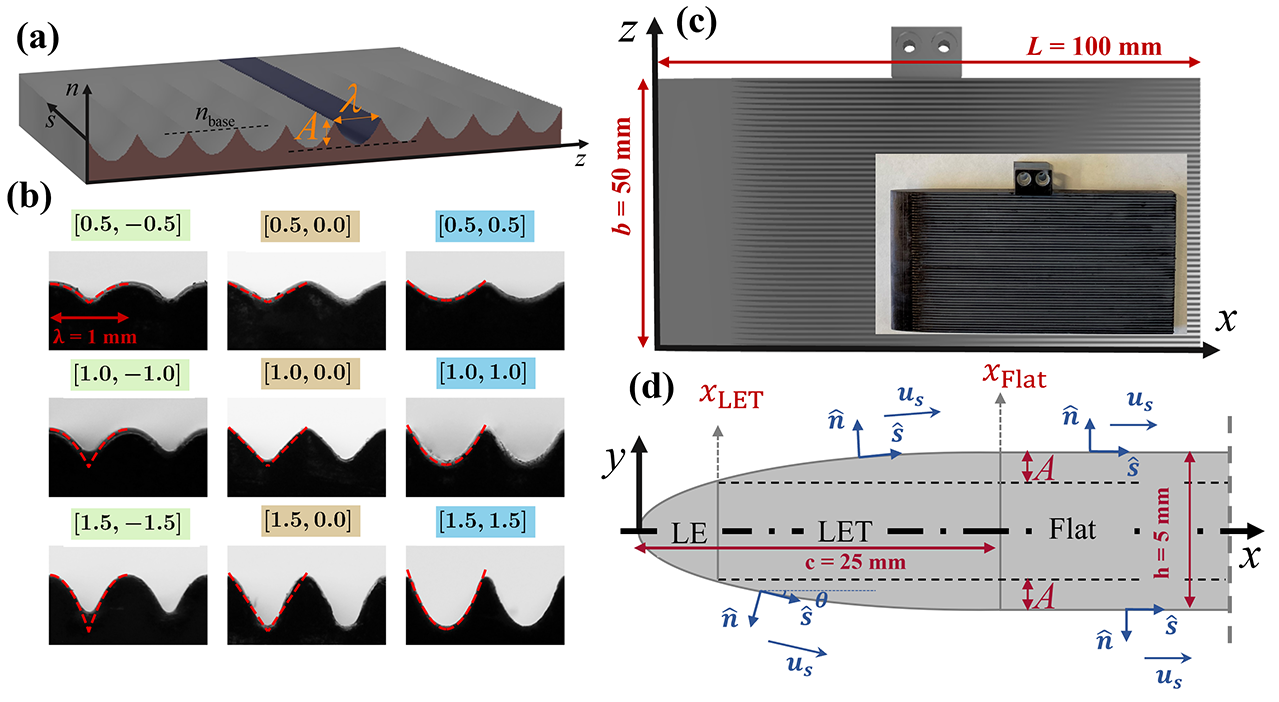}
    \caption{(a) Schematic of a riblet surface with spacing $\lambda$ and height $A$ and a concave cross-sectional profile. (b) Images of the cross-sectional profiles of riblets samples. For all the samples $\lambda = 1$ mm, and the respective [${\cal R}, \kappa_2$] values are listed above each sample. (c) Schematic of the textured sample design and the front view of an actual sample. (d) Schematic of the side view of the sample, including the leading edge of the textured samples and the early part of the Flat region. }
    \label{fig:riblets}
\end{figure}

For symmetric and periodic riblets, the shape of the cross-sectional profile of the riblets can be defined only for half of each unit which is mirrored and then repeated as needed. Here, we focus on the simplest family of curved riblets defined using a second-order polynomial \citep{raayai2022polynomial}, where for a unit element with spacing $\lambda$ and height $A$ (figure \ref{fig:riblets}(a)),

\begin{equation}\label{eq:profile}
    \frac{n_{\rm w}}{\lambda/2} = \kappa_2\left(\frac{z}{\lambda/2}\right)^2+(-\mathcal{R}-\kappa_2)\left(\frac{z}{\lambda/2}\right)+ \dfrac{n_{\rm base}}{\lambda/2}  \ \ \ \ \ \ \  0\leqslant z \leqslant (\dfrac{\lambda}{2})
\end{equation}

\noindent where $n_{\rm w}$ is the cross-sectional profile of the riblet wall in the normal direction, $z$ is the spanwise coordinate, $n_{\rm base}$ is the height of the surface that the textures are carved from (equal to the height of the peaks of the textures), ${\cal R} = A/(\lambda/2)$ is the height-to-half-spacing ratio as a measure of sharpness \citep{raayai2022polynomial, raayai2017drag, raayai2019geometric, raayai2020geometry}, and $\kappa_2$ is a dimensionless curvature parameter \citep{raayai2022polynomial}. Within this framework, $\vert \kappa_2 \vert \leqslant {\cal R}$, where $\kappa_2<0$ corresponds to convex textures, $\kappa_2>0$ corresponds to concave textures, and $\kappa_2 = 0$ places the commonly used triangular (V-groove) textures as a subset of this riblet family. We focus on the case of textures with $\lambda = 1$ mm and three height-to-half-spacing ratios of 0.5, 1.0, and 1.5, and one concave ($\kappa_2 = {\cal R}$), one convex ($\kappa_2 = -{\cal R}$), and one triangular case within each family (9 different textures, see figure \ref{fig:riblets}(b)). Throughout this paper, the samples are identified using the vector $\boldsymbol{\kappa} = [ {\cal R}, \kappa_2]$.

Here, we use the variable $s$ and $n$ as the local coordinate system, tangential and normal to the surface (figure \ref{fig:riblets}(d)), and use the variable $\lambda$ instead of the commonly used $s$ for spacing of the riblets. In addition, we reserve the variable $t$ for time and use $h$ for the thickness of the finite-sized samples and employ $A$ instead of $h$ for the height of the riblet units. These simple changes compared to the commonly used variables allow us to avoid confusion. In addition, we use the word ``smooth'' as the opposite of a riblet surface while we will use the word ``flat'' as opposed to a ``curved'' surface.

\subsection{Textured Samples} \label{sec:sample}

We use fully textured samples where the riblets are fabricated seamlessly with the samples. The base of the samples has the form of a symmetric slender plate, comprised of a leading edge, and a main body (100 mm in length (L), 50 mm in width (b), and 5 mm in depth (h)). The leading edge is streamlined in the form of an ellipse (similar to the design used by \citet{fu2023doublelightsheet}) up to $x=25 \ {\rm mm}$ where it asymptotically meets a flat main body which extends to a blunt trailing edge ($25 \ {\rm mm} \leqslant x \leqslant 100 \ {\rm mm}$ (see figure \ref{fig:riblets}(c)). The riblets are carved out of the base geometry on both sides of the sample. In this case, with the coordinate system shown in figures \ref{fig:riblets}(c-d), the base height in the main (Flat) region of the body,  $\vert y_{\rm base} \vert$, is at half of the thickness, $h/2$, on either side. The peaks of the textures reside at this height on either side and the troughs are located at $\vert y_{\rm base} \vert - A$.  However, this is only true for the Flat region of the sample within ($25 \ {\rm mm} \leqslant x \leqslant 100 \ {\rm mm}$) and prior to that, due to the curved nature of the leading edge, the textures appear where there is enough thickness (see figure \ref{fig:riblets}(c-d)) and grow to their maximum height $A$ at the end of the elliptical leading edge. We call this region, ``leading edge textured'', LET, and the non-riblet segment of the leading edge as LE. At the leading edge of fully riblet-covered flat plates, for at least $x/\lambda<9$, riblets experience larger spanwise-averaged wall shear (normalised with $\lambda$) compared with the smooth reference \citep{raayai2017drag}, and to mitigate that, we mimic the distribution of the denticles on the nose of sharks \citep{lauder2016structure} (which start from smooth in the nose area and ribs appear later along the body) in the design of the streamlined leading edge of the samples. Thus, instead of enforcing a constant height-to-half-spacing ratio for the textures, we keep the wavelength constant and let the height of the textures grow from zero to the final height, $A$, at the end of the elliptic area and let the BL evolve with the evolution of the riblets.

Riblet samples (and one smooth for comparison) are 3D printed (Formlabs Form3 3D printer and a photo-polymer resin, figures \ref{fig:riblets}(b-c)). After printing, the spacing and the height of the riblets are measured using Bruker ContourX-500 Optical Profilometer and analyzed with the open-source software package Gwyddion. The measurements are conducted at 4 different random locations on each side of the samples, with each location covering about $2.8\ {\rm mm}^2$ of the projected area and containing at least one riblet unit, and the mean of the measurements and their 95\% confidence intervals are reported in table. \ref{tbl:geom}. Due to the limited resolution of the 3D printer, the final height of the riblets are smaller than the design heights and result in lower apparent $\cal R$ values for the samples, but do not affect the performance of the riblets. Throughout the paper the samples are identified with their design names as listed in the first column of table \ref{tbl:geom}. 

\begin{table}
\begin{center}
\def~{\hphantom{0}}
    \begin{tabular}{c|cccccc} 
  Sample Name   &  ${\cal R}$ & $\kappa_2$ & Height, $A$ & Spacing, $\lambda$ & Texture Start, $x_{\rm LET}$ & $\alpha$ \\
   & [-] & [-] & [$\upmu$m] & [$\upmu$m] & [mm]  & [$^\circ$]\\
  Smooth         &    0.0   &  0.0  &  NA  &   NA & NA  & $0.49^{\circ} \pm 0.05^{\circ}$\\
  $[0.5, -0.5]$          &    0.5    &  -0.5  &  222 $\pm$ 3  & 1002 $\pm$ 5 &   (14.10) 15.00$\pm$ 0.41 & 1.07$^{\circ}$ $\pm$ 0.01$^{\circ}$\\
 $[0.5, \ \ 0.0]$ &    0.5    &  0.0  &   215 $\pm$ 6  &  995 $\pm$ 9 & (14.10) 15.27$\pm$0.54  & 1.32$^{\circ}$ $\pm$ 0.02$^{\circ}$\\
 $[0.5, \ \ 0.5]$    &    0.5    &  0.5  &  200 $\pm$ 3  &  998$\pm$8 &   (14.10) 14.25$\pm$0.25  & 1.07$^{\circ}$ $\pm$ 0.01$^{\circ}$\\
 $[1.0, -1.0]$          &    1    &  -1.0  &  303 $\pm$ 2  & 1004 $\pm$10 &  (10.00) 10.79 $\pm$ 0.30 & 1.07$^{\circ}$ $\pm$ 0.01$^{\circ}$\\
 $[1.0, \ \ 0.0]$ &    1   &  0.0  &  372 $\pm$ 6  & 994 $\pm$ 6 &  (10.00) 10.74$\pm$ 0.21 & 1.14$^{\circ}$ $\pm$ 0.02 $^{\circ}$\\
 $[1.0, \ \ 1.0]$    &    1    &  1.0  &  405 $\pm$ 5  &  997 $\pm$ 4 &   (10.00) 9.7 $\pm$ 0.10   & 1.21$^{\circ}$ $\pm$ 0.01 $^{\circ}$\\
 $[1.5, -1.5]$          &    1.5    &  -1.5  &  562 $\pm$ 4  &  998 $\pm$6 &  (7.15) 8.18 $\pm$ 0.32 & 1.48$^{\circ}$ $\pm$ 0.01$^{\circ}$\\
 $[1.5, \ \ 0.0]$ &    1.5    &  0.0  &  636 $\pm$ 6  & 1005$\pm$9 &  (7.15) 7.27 $\pm$ 0.07 & 0.95$^{\circ}$ $\pm$ 0.02$^{\circ}$\\
 $[1.5, \ \ 1.5]$    &    1.5    &  1.5  &  644 $\pm$ 2  &  986$\pm$ 5 &  (7.15) 7.15 $\pm$ 0.17  & 1.59$^{\circ}$ $\pm$ 0.02$^{\circ}$\\
\end{tabular} 
    \caption{Details of the geometry of the riblets and setup. The locations of the starts of the textures ($x_{\rm LET}$) are measured using a caliper (design values in the parenthesis).}
    \label{tbl:geom}
\end{center}
\end{table}

\subsection{Experimental Procedure}
\begin{figure}
    \centering
    \includegraphics[width=1\linewidth]{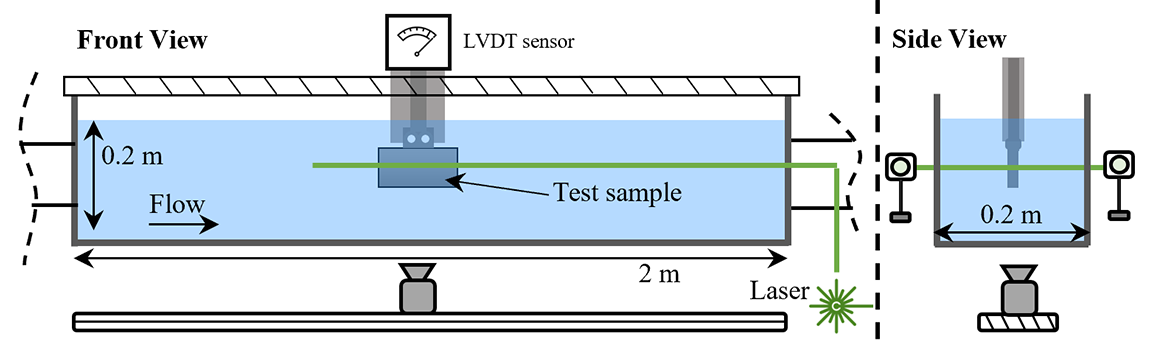}
    \caption{Schematic of the experimental setup, showing front and side views of the water tunnel, the installed sample, and the PIV setup.}
    \label{fig:exp_setup}
\end{figure}

Experiments are performed following the procedure described previously \citep{fu2023doublelightsheet, fu2023multisheet}. In summary, the samples described in section \ref{sec:sample} are attached to a dynamometer consisting of linear variable differential transformers (to measure the total drag force) and suspended in a water tunnel (cross-sectional area of $20 \ {\rm cm} \times 20 \ {\rm cm}$ and $2 \ {\rm m}$ in length, see figure \ref{fig:exp_setup}). The samples are positioned at around $75 \ {\rm cm}$ from the tunnel entrance and close to the middle of the cross-section to reduce the effects of the tunnel walls on the measurements. The experiments are performed at three free stream velocities less than $0.25 \ \rm{m/s}$ ($0.122$, $0.185$, and $0.242 \ {\rm m/s}$) corresponding to global Reynolds numbers, ${\rm Re}_L = \rho U_{\infty}L/\mu$, of 12,200, 18,500, and 24,200 (turbulence intensity of the free stream is less than or about $1\%$).

The velocity field all around the sample is captured using a double-pulsed planar (two-dimensional, two-component, 2D-2C) PIV and double-illumination, consecutive-overlapping procedure \citep{fu2023doublelightsheet, fu2023multisheet}. The setup is comprised of a double-pulsed Nd:YAG laser operated at $15 \ {\rm Hz}$ repetition rate and the nominal output energies of $10 \ {\rm mJ}$ or $20 \ {\rm mJ}$ per pulse for different free-stream velocities. The high-speed camera is set to a resolution of 720 $\times$ 1920 pixels and the timing between the two consecutive pulses is adjusted to $\delta t =$ 1000 $\upmu$s for free-stream velocities of 0.122 and 0.185 m/s, and $\delta t =$ 900 $\upmu$s for 0.242 m/s to allow for the slowest particles to have visible movement and the fastest particles to not move beyond half of the PIV analysis window (set at 32 pixels). 

To get simultaneous access to both sides of a non-transparent sample and avoid any assumptions regarding the symmetry of the flow around the foils, we use a double-illumination strategy \citep{fu2023doublelightsheet, fu2023multisheet}. This optical setup  is comprised of a half-wave plate and a polarizing beam splitter to divide the incoming laser into two almost equal beams. These beams are then guided toward either side of the sample through additional mirrors where two combinations of light-sheet optics, involving two spherical lenses and one cylindrical lens are positioned on opposite sides of the water tunnel to create  light sheets with thicknesses of about $ 1 \ {\rm mm}$. 

To have access to the velocity field in both the BL and the far field, we use a consecutive-overlapping imaging procedure \citep{fu2023doublelightsheet, fu2023multisheet}. We attach the camera to a computer numerically controlled rail (CNC) and traverse the entire area of interest of the flow with about 40-45\% overlap between the fields of views. To have access to the velocity profiles in the BL, we use a magnification of 1  px $\equiv$ 15-16 $\upmu {\rm m}$  which limits the total field of view of the camera at a time and the consecutive-overlapping imaging allows us to overcome this limitation and capture a larger area of interest. Here we use a one-dimensional sweep in the streamwise direction, with 36 overlapping steps to capture a span of about 180 $\rm mm$ in the streamwise direction and about 25 mm in the normal direction. On either side, the BL thickness does not go beyond 3 ${\rm mm}$. 

At each step, 50 image pairs are captured. \citet{fu2023doublelightsheet} have previously shown that 50 image pairs are enough for the measurements to reach convergence in a similar range of Reynolds number as that used here
. The images (and subsequently the analyzed velocity fields) are then stitched together based on the global locations of the camera as controlled with the CNC rail. The PIV images at each step of imaging are analyzed using an in-house Python script utilizing the open-source software OpenPIV \citep{alex_liberzon_2020_3911373} (with $32 \times 32$ windows and a search area of $64 \times 64$ with $85\%$ overlap) to find the velocity in the $x$ and $y$ directions ($u$, and $v$ respectively) with an additional correction loop for the velocities close to the wall region \citep{fu2023doublelightsheet} to avoid bias errors due to the large shear rates close to the walls \citep{kahler2012uncertainty}. 

\subsection{Data Analysis}

\subsubsection{Angle of Attack}

While each sample is positioned as close to parallel to the streamwise direction as possible, there is always the potential for a slight non-zero angle of attack, $\alpha$, not visible to the naked eye in the setup and the final angle of attack of the sample with respect to the free-stream velocity is not known \textit{a priori}. This $\alpha$ results in an asymmetry in the flow and subsequently differences between the local responses of the Front ($y>0$) and Back ($y<0$) sides of the samples. To find this angle of attack more accurately for each of the experiments, we use the velocity field upstream of the leading edge and a potential model of flow past an elliptical leading edge (previously described by \citet{fu2023doublelightsheet}) and fit the velocity measurements to the potential model to find the $\alpha$ between the sample and the free-stream velocity as listed in the last column of table \ref{tbl:geom}. Since the experiments with each sample were performed in one session (only changing the velocity), the values of the $\alpha$ are a mean of fits at 30 locations within $x/L<-0.14$ of the leading edge and for all the 3 runs (different ${\rm Re}_L$) of each sample for a total of 90 values. Due to the angle of attack, and with the geometric configuration of the experimental setup, the Front/Back sides of the samples coincide with the suction/pressure sides of the foils respectively. 

\subsubsection{Shear Stress Distribution} \label{shearmethod}

To calculate the local shear stress distribution, we employ the PIV data, to find the velocity profiles at each $x$ location and calculate the local velocity gradient normal to the wall as a function of the local Reynolds number, ${\rm Re}_x = \rho U(x) x/\mu$, where $U(x)$ is the velocity at the edge of the BL. Numerical simulations of BL over riblet surfaces \citep{goldstein1995direct, choi1993direct, raayai2017drag, raayai2020geometry} have previously shown that in presence of the riblets, the velocity profiles are dependent on all three dimensions, $\mathbf{u}(x,y,z)$, and dependent on the spanwise direction, $z$, in a periodic manner. This is directly a result of the shape of the riblet surface and the no-slip wall. However, the previous numerical and experimental observations \citep{raayai2020geometry, raayai2019geometric, raayai2017drag, choi1993direct, djenidi1989numerical, djenidi1994laminar, djenidi1996laser} all show that the spatial variations in the velocity profiles are dominant inside the riblet unit element near the wall while moving outside the grooves all velocity profiles at a given $x$ location, collapse unto each other with no spanwise dependence visible. 

In planar PIV, the flow measurements performed in the thin plane of the light sheet, are an average of the velocity field within this light sheet. Thus, in our current experiments, we set the thickness of the light sheet to about 1 mm to match the spacing of one riblet unit and ensure the depth of field of the camera is also around 1 mm to capture a spanwise-averaged velocity measurement. Writing the velocity profile as $\mathbf{u}(x,y,z)$, the measured velocity profile is then $\langle \mathbf{u} \rangle (x,y)$ where $\langle \cdots \rangle (x,y)$ denotes the spanwise-averaging operation. Due to the opaqueness of the samples, only the velocity distribution outside of the grooves ($n>0$) is visible to the camera, and the rest of the profiles have their respective no-slip wall hidden inside the grooves, below the level accessible to imaging. Thus, the origin of the average velocity $\langle \mathbf{u} \rangle$ is located below the peak of the riblets ($n<0$). We use this origin as a representative average origin of the velocity within one texture unit and denote it as the ``effective origin'' of the riblets, $n_0$. 

From the PIV data, we calculate the spanwise-averaged velocity gradient and subsequently the local spanwise-averaged shear stress, $\langle \tau \rangle = \mu \partial \langle u_s \rangle /\partial n$. Instead of depending on only a few measurement points close to the wall, we use as many of the measured velocity points as possible and characterize the velocity profiles in a mathematical format and fit the profiles to an appropriate functional form. Here, we employ an updated form of the Falkner-Skan (FS) \citep{falkner1931lxxxv} family of BL in a localized manner to capture the behaviour of the tangential velocity, denoted by $\langle u_s \rangle = \langle u \rangle \cos \theta + \langle v \rangle \sin \theta$ (where $\theta$ is the local angle between $\hat{s}$ and $x$ directions). In this updated formulation, we define $\langle u_s \rangle$ as a function of the local Reynolds number, ${\rm Re}_{x} = {\rho x U(x)}/{\mu}$, $n$, $m$, and the effective origin $n_0$, in the form of $\langle u_s \rangle = \mathcal{H}({\rm Re}_{x},n;m,n_{0})$, where the averaged velocity profile is a function of an updated similarity variable $\eta^*$, of the form ${\langle u_s\rangle}/{U} = \mathcal{F}'(\eta^*)$ with $\eta^*$ defined as

\begin{equation}\label{eq:eta}
    \eta^* = \dfrac{(n-n_{0})}{x} \sqrt{{\rm Re}_x\left(\dfrac{m+1}{2}\right)}.
\end{equation}

\noindent Now, with this formulation, and the experimental data for $\langle u_s \rangle$, the local normal direction, $n$, and ${\rm Re}_x$, we \emph{locally} fit the data to the FS solutions to the BL equations and find the best values of $m$ and $n_0$ that capture the profiles at every location. A few examples of the fitted velocities along the suction side of the [1.0, 0.0] sample operated at ${\rm Re}_L = 18,500$ are shown in figure \ref{fig:Vel_CV}(a) in appendix \ref{appA}. In the curved leading edge area (LE and LET), $n \nparallel y $, and for simplicity instead of $s$ we use $x$ and ${\rm Re}_x$ to characterize the streamwise velocity profiles (with every $x$ having a unique mapping to $s$ and vice versa, we let the marginal effect of the difference between the magnitude of $s$ and $x$ to be captured in the $m$ parameter). The first reliable fit is around $x=1 \ {\rm mm}$ where $s/x \approx 1.27$ and by the end of the LET, $s/x \approx 1 .015$. In the Flat region, the streamwise and normal direction align with $x$ and $y$ coordinates ($\langle u_s \rangle = \langle u \rangle)$. 

Note that due to the finite thickness of the sample, the BL edge velocity, $U(x) \geqslant U_{\infty}$, and thus local ${\rm Re}_x$ values are larger than Reynolds numbers calculated using $\rho U_{\infty}L/\mu$ and thus ${\rm Re}_{x=L}> {\rm Re}_L$. Here, we do not find one single $m$ for the entire flow, but use this family of FS solutions and the parameter $m$ as mathematical tools to characterize the local behaviour of the flow field, especially including terms that cannot be captured directly with the planar PIV measurements (discussed further in the upcoming section \ref{fittingParameterDist}).

Knowing the mathematical form of the FS solutions as well as the distribution of the $m$ and $n_0$, the local spanwise-averaged shear stress distribution along each side of the plate is

\begin{equation}\label{eq:tau}
   \langle \tau_{\rm w} \rangle (x) = \mu \left. \dfrac{\partial  \langle u_s \rangle }{\partial n} \right \vert_{n=0} = \left. \left(\frac{m+1}{2}\right)^{0.5}\frac{\rho U(x)^2}{\sqrt{{\rm Re}_x}}\mathcal{F}''\right \vert_{\eta^*=\eta^*_0}
\end{equation}

\noindent and the spanwise-averaged skin friction coefficient is determined by $\langle C_f \rangle(x) = {  \langle \tau_{\rm w} \rangle (x)  }/{({1}/{2})\rho U(x)^2}$. As written in equation \eqref{eq:tau}, we use the gradient of $\langle u_s \rangle$ profiles on the $n=0$ plane to find $\langle \tau_{\rm w} \rangle$ distribution at every location. We show that using a simple control volume analysis inside the grooves (as discussed in the appendix \ref{appA}), the gradient of $\langle u_s \rangle$ profiles on the $n=0$ is able to capture the essence of the velocity gradient distribution at the riblet wall while also capturing the effect of the excess wetted surface area of the riblets compared with the smooth reference. Note, at each local Reynolds number, the direct effect of $n_0$ on the magnitude of the $\langle \tau_{\rm w} \rangle$, as written in equation \eqref{eq:tau}, is mainly hidden in the value of the $\eta^*_0$ at the peak of the grooves. We use the $\langle \tau_{\rm w} \rangle$ of the riblets and the $\tau_{_{0}}$ of the smooth reference as local measures for comparing the frictional (shear) response of the surfaces. 

\subsubsection{Pressure and pressure gradient distribution} \label{sec:pressurecalc}

We use the PIV data to find the pressure gradient and pressure distribution by using the Reynolds averaged Navier Stokes (RANS) equations in $x$ and $y$ direction and line-integration of the gradient terms in those directions \citep{fu2023doublelightsheet, suchandra2023impact,Liu_2006, Charonko_2010, Kat_2013, Oudh_2013, Liu_2016, Liu_2020, Nie_2022}

\begin{equation} 
\label{eq:dpdx}
    \dfrac{\partial p}{\partial x} = \biggl[ -\rho \left(u \frac{\partial u}{\partial x} + v \frac{\partial u}{\partial y} \right)
+ \mu \left( \frac{\partial^2 u}{\partial x^2} + \frac{\partial^2 u}{\partial y^2} \right) - \rho \left(\frac{\partial \overline{u' u'}}{\partial x} + \frac{\partial \overline{u' v'}}{\partial y} \right) \biggl]
\end{equation}

\begin{equation}\label{eq:py}
\dfrac{\partial p}{\partial y} = \biggl[ -\rho \left(u \frac{\partial v}{\partial x} + v \frac{\partial v}{\partial y} \right)
+ \mu \left( \frac{\partial^2 v}{\partial x^2} + \frac{\partial^2 v}{\partial y^2} \right) - \rho \left(\frac{\partial \overline{u' v'}}{\partial x} + \frac{\partial \overline{v' v'}}{\partial y} \right) \biggl] 
\end{equation}

\noindent where $\rho$ and $\mu$ are the fluid's density and dynamic viscosity respectively and $\overline{(...)}$ denotes ensemble-averaging with the instantaneous velocity vector $\mathbf{\tilde{u}}(x,y,t)$ written as the sum of the mean and the fluctuating component, $\mathbf{\tilde{u}}(x,y;t) = \mathbf{u}(x,y) + \mathbf{u'}(x,y;t)$. We only use the above equations far enough from the riblets that we can assume the 3D effects of the riblets have subsided and $\mathbf{u}(x,y,z) = \mathbf{u}(x,y) = \langle \mathbf{u} \rangle (x,y) $ and $p(x,y,z) = p(x,y) = \langle p \rangle (x,y)$. Even though the BL stays laminar all over the body, the wake of the sample becomes turbulent and for completeness, we include the Reynolds stresses in the RANS equations as shown. We set the $p_{\rm ref}$ at the furthest distance prior to the leading edge and design combinations of horizontal and vertical linear paths from the reference point to a point of interest and use integration either in $x$ or $y$ direction to find the pressure at the point of interest.

\section{Results and discussion} \label{sec:results}

\subsection{Drag} \label{sec:drag}

\begin{figure}
    \centering
    \includegraphics[width = 1 \textwidth]{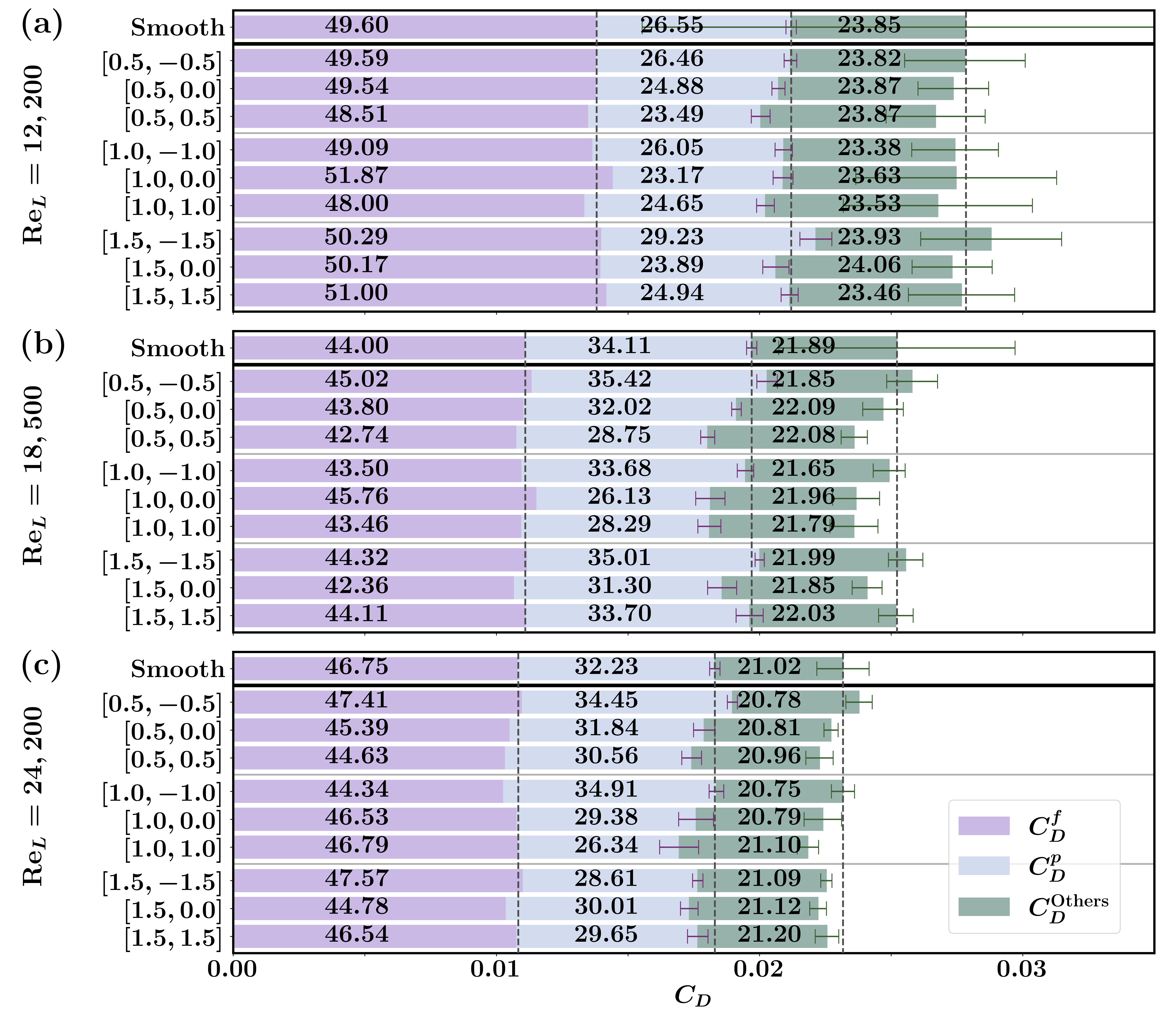}
    \caption{Decomposition of the total drag force, in terms of the drag coefficient, into friction, $C_D^{f}$, pressure, $C_D^{p}$, and $C_D^{\rm Others}$ components for experiments performed at global Reynolds numbers (a) ${\rm Re}_L = 12,200$, (b) ${\rm Re}_L = 18,500$, and (c) ${\rm Re}_L = 24,200$ for all samples. Percentage of the contribution of each component with respect to the total drag of the smooth reference sample is presented on the bars (the sum of the values for smooth samples comes to 100\%). }
    \label{fig:CfCpC3D}
\end{figure}

First, we present the total drag force experienced by each of the samples and the reference smooth case at ${\rm Re}_L = 12,200$, $18,500$, and $24,200$ in figure \ref{fig:CfCpC3D}. Here, using a 3-tiered measurement approach \citep{fu2023doublelightsheet}, we linearly decompose the total drag measured via the dynamometer into three components; $D_{f}$, the frictional drag force, $D_{p}$, the pressure drag, and $D_{\rm others}$, which is all the 3D and other effects that are not captured by the previous two components ($D = D_{f} + D_{p} +  D_{\rm others}$). All the drag components are normalized by $(1/2)\rho U_{\infty}^2 2Lb$, where $b$ is the width of the sample in the spanwise direction, and presented in the form of drag coefficients, $C_D^f$, $C_D^p$, and $C_D^{\rm others}$. On the first tier, the friction drag, due to the shear stress distribution, is found using the integral of the $\langle \tau_{\rm w} \rangle$ distribution on both sides of the sample and discussed more in sections  \ref{sec:friction} and \ref{sec:shear}. On the second tier, we find the pressure drag cumulatively using a control volume analysis as discussed later in section \ref{sec:pressureDrag}. On the last level, the $D_{\rm others}$ then is found by subtracting the pressure and fiction drags from the dynamometer measurements. 

As seen in figure \ref{fig:CfCpC3D}, within families of ${\cal R} = 0.5$ and ${\cal R} = 1.0$, concave textures with $\kappa = {\cal R}$ exhibit the lowest drag force compared to the rest of the samples and convex ones with $\kappa = -{\cal R}$ the largest drag. Triangular textures with $\kappa = 0$ show drag values in between that of the $\kappa = \pm {\cal R}$ cases, sometimes close to one or the other depending on the case. However, going to the sharper textures of ${\cal R} = 1.5$, at all Reynolds numbers, sample [1.5, 0.0] (triangular) remains the lowest drag, with the concave [1.5, 1.5] riblet in the second place and convex [1.5, -1.5] seeing the largest drag for ${\rm Re}_L = 12,200$, and $18.500$. At ${\rm Re}_L = 24,200$ the the concave and convex samples, [1.5, 1.5] and [1.5, -1.5], experience nearly similar values of drag.

This trend in the drag as a function of the $\kappa_2$, is similar to the pattern previously reported in in the Couette flow and early Taylor vortex regime in a Taylor-Couette (TC) flow {\mbox{\citep{raayai2022polynomial}}} where for shallow textures with ${\cal R} \leq 1$, torque of the riblet-covered rotors decreases as $\kappa_2$ is increased (at a constant ${\cal R}$). Going to sharper textures such as $\kappa_2 = 1.5$, the torque does not monotonically follow the $\kappa_2$ parameter and the lowest recorded torque is around  $0 \leq \kappa_2 \leq 0.5$ for riblets with spacing of 1 mm (spacing normalized by the gap of the TC cell of 0.13). In addition, previous experiments of \citet{walsh1984optimization}, show larger drag reductions for concave up semi-circular riblets compared with triangular textures and nearly no drag reduction for convex-up semi-circular riblets. They also show that increase in the radius of curvature at the trough of triangular textures can reduce the drag while an increase in the radius of curvature at the peak of the grooves slowly increases the drag experienced by the riblets. 
 
In comparison to the smooth reference, the majority of the cases tested here show some level of drag reduction. For ${\cal R} = 0.5$ family, only the [0.5, 0.5] and [0.5, 0.0] samples were drag-reducing in all cases. In the ${\cal R} = 1.0$ family, the [1.0, 1.0] sample experiences the largest drag reduction among the rest of the family, with [1.0, 0.0] seeing slightly lower reductions, and the [1.0, -1.0] samples experiencing no reduction at ${\rm Re}_L = 24,200$ or reductions of slightly larger than 1\% for the lower Reynolds numbers. For the ${\cal R} = 1.5$ family, the [1.5, 0.0] is able to stay drag-reducing for all the tested Reynolds numbers, while for ${\rm Re} = 12,200$ and $18,500$ the [1.5, -1.5] is fully drag-increasing, and [1.5, 1.5] stays nearly neutral. At the largest Reynolds number of 24,200, all ${\cal R} = 1.5$ samples become drag-reducing.

Breaking down the measured drag into components, we see that the friction and the pressure drags do not follow the same trend as that of the total drag. Except for a few of the cases reported, the total frictional component of the drag force only experiences marginal changes due to the presence of the textures as discussed more in section \ref{sec:friction}. However, we record substantial levels of reductions in the pressure drag for quite a number of the samples which help in reducing the drag even when the frictional drag is unchanged. While the effect of riblets in changing the pressure distribution and pressure drag has largely been unexplored, recent numerical simulations of riblet covered bodies \citep{mele2016performance, mele2020effect} have also confirmed the ability of riblets to impact the pressure drag as discussed in section \ref{sec:pressureDrag}.

At the lowest Reynolds number, ${\rm Re}_L = 12,200$, and among the family of ${\cal R} = 0.5$, only the $D_{f}$ of the [0.5, 0.5] sample experiences a reduction of 2.2\% compared to the smooth reference, contributing 1\% to the total of 4.1\% reduction in the total drag. Among the ${\cal R} = 1.0$ family, both the [1.0, -1.0] and [1.0, 1.0] samples experience 1\% and 3.2\% reduction in the $D_{f}$, contributing to 0.5\% (out of 1.48\%) and 1.6\% (out of the 3.82\%) to the total drag reductions of these samples respectively. The [1.0, 0.0] sample on the other hand, experiences a 4.6\% increase in the frictional drag, requiring the pressure drag to contribute a 3.4\% reduction to overcome the added 2.27\% for the sample to experience a 1.1\% reduction in the total drag. Similar trends can be extracted for the samples at larger Reynolds numbers (see figure \ref{fig:CfCpC3D}).

Overall, $C_D^{f}$ of ${\cal R} = 0.5$ and  ${\cal R} = 1.5$ follows a similar trend as that of the total drag force, while for the ${\cal R} = 1.0$ family, unlike the total drag, the trend in $C_D^{f}$ is non-monotonic with respect to $\kappa_2$, with the [1.0, 1.0] sample experiencing the lowest and [1.0, 0.0] experiencing the highest $C_D^{f}$ at ${\rm, Re}_L = 12,200$ and $18,500$, and at the largest ${\rm Re}_L = 24,200$, [1.0, 1.0] and [1.0, 0.0] having nearly similar value of $C_D^{f}$ which is larger than that of the [1.0, -1.0]. Among the ${\cal R}= 0.5$ case, the magnitude of the pressure drag also takes a decreasing trend with the $\kappa_2$ parameter, with [0.5, 0.5] experiencing the lowest $C_D^{p}$ and [0.5, -0.5] experiencing the largest one. As a result, [0.5, 0.0] and [0.5, 0.0] also experience a reduction in $C_D^{p}$ compared with the smooth sample while [0.5, -0.5] stays either neutral or $C_D^{p}$-increasing. Thus, within this family, cumulatively with $C_D^{f}$, only [0.5, 0.0] and [0.5, 0.5] are drag-reducing. Among the ${\cal R} = 1.0$ family, at ${\rm Re}_L = 12,200$ and $18,500$ the [1.0, 0.0] sample experiences the lowest $C_D^{p}$ while at ${\rm Re}_L = 24,200$ the [1.0, 1.0] sample has the lowest $C_D^{p}$ and [1.0, -1.0] sample experiencing the largest $C_D^{p}$ among all. The [1.0, 0.0] sample which could not offer any frictional reduction, experiences a large enough reduction in the pressure drag to be cumulatively drag-reducing. Similarly, for the [1.0, 1.0] sample, the reduction achievable in the pressure drag enhances the cumulative drag reduction of this sample. For the [1.0, -1.0] the pressure drag is not able to enhance the drag reduction at ${\rm Re}_L = 12,200$ and $18,500$ and at $24,200$ it even diminishes the drag reduction that was achieved from the frictional component. For the sharpest riblet family, at ${\rm Re}_L = 12,200$ and $18,500$, [1.5, 0.0] experiences the lowest pressure drag and [1.5, -1.5] the largest, and only [1.5, 0.0] is able to see a cumulative reduction in the drag force. At ${\rm Re}_L = 24,200$, the pressure drag experienced by the three samples are nearly similar and all three are able to capture a reduction in the pressure drag which also leads to them being cumulatively drag-reducing as well.  

Lastly, as it can be seen from the percentages listed on the bars of figure \ref{fig:CfCpC3D}, the contributions attributed to the $D_{\rm others}$, not captured via the PIV measurements, do not show any clear dependence on the textures and have nearly similar values and is not discussed further.  

\subsection{Friction Drag} \label{sec:friction}

Out of the total drag force exerted on the slender samples, between 40-50\% of the drag is due to the frictional component, $D_{f}$ (figure \ref{fig:CfCpC3D}, purple portion of the bar plots). This component is comprised of the frictional drag experienced on either side of the sample found using the integral of the spanwise-averaged shear stress distributions as 

\begin{equation}
    D_{f} = D_{_{\rm Front}} + D_{_{\rm Back}} = b\left(\int_0^{{L}} \langle \tau_{_{\rm w}} \rangle_{_{\rm Front}} dx + \int_0^{L} \langle \tau_{_{\rm w}} \rangle_{_{\rm Back}}dx \right).
\end{equation}

\noindent and along each side, it can be divided into the contributions of the LE, LET, and the Flat regions as written in the form of

\begin{equation}
\begin{split}
    D_{_{\rm Front}} &= b \left(\int_0^{x_{_{\rm LET}}} \langle \tau_{_{\rm w}} \rangle_{_{\rm Front}}(x) dx + \int_{x_{_{\rm LET}}}^{x_{_{\rm Flat}}} \langle \tau_{_{\rm w}} \rangle_{_{\rm Front}}(x) dx + \int_{x_{_{\rm Flat}}}^{L} \langle \tau_{_{\rm w}} \rangle_{_{\rm Front}}(x) dx \right)
\end{split}
\label{contributions}
\end{equation}

\noindent and $D_{_{\rm Back}} = D_{_{\rm Back}}^{_{\rm LE}} + D_{_{\rm Back}}^{_{\rm LET}} + D_{_{\rm Back}}^{_{\rm Flat}} $. Note that the spanwise-averaged wall shear stress distribution, $\langle \tau_{_{\rm w}} \rangle$, is the shear stress component exerted along the surface, or $\hat{s}$, with $\theta$ the local angle between the $x$ and $\hat{s}$ directions. Thus, with $dx = \cos \theta ds$, we have $\langle \tau_{\rm w} \rangle \cos \theta ds = \langle \tau_{_{\rm w}} \rangle dx$.

\begin{figure}
    \centering
    \includegraphics[width = \textwidth]{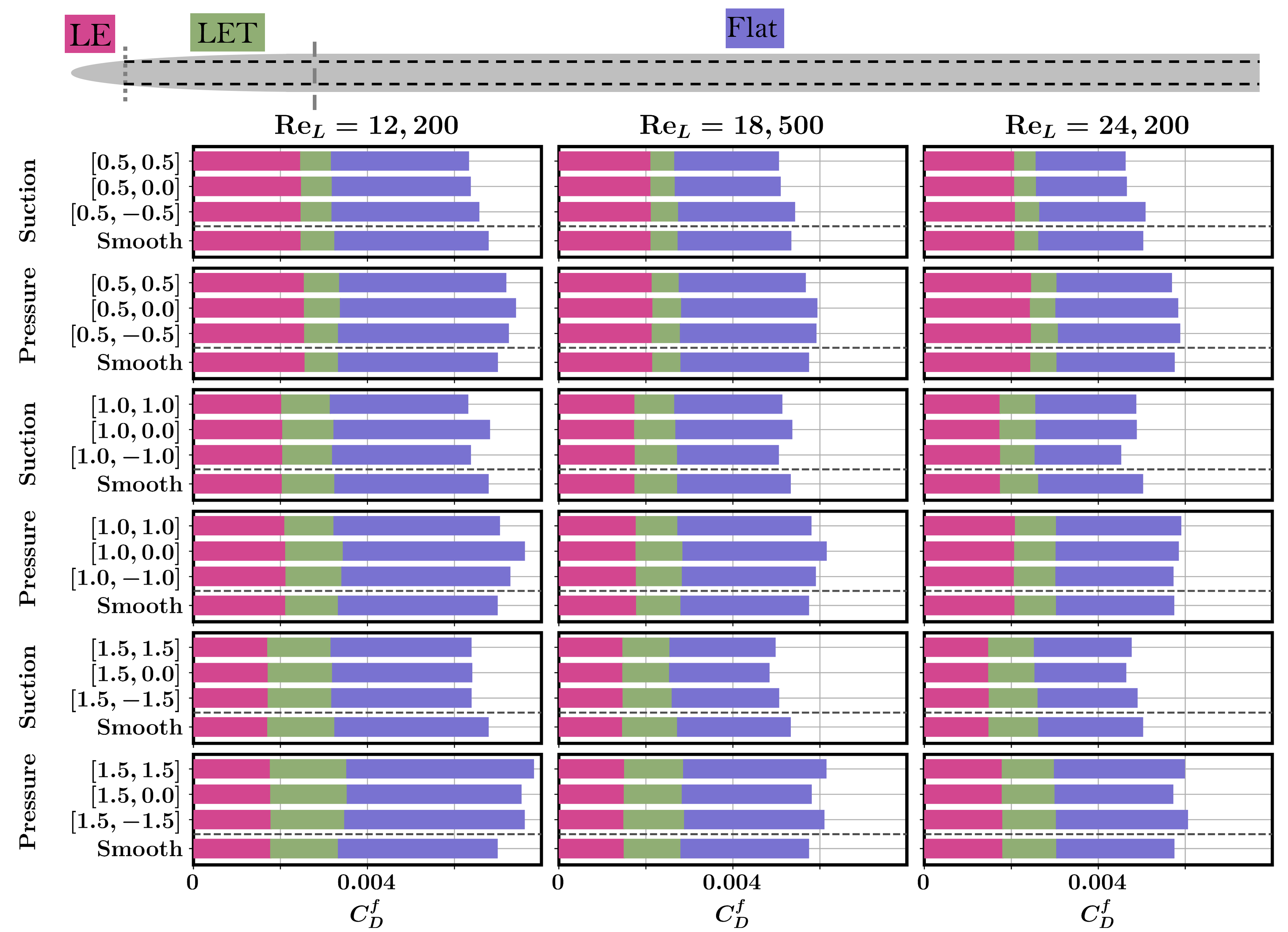}
    \caption{Bar plots showing the contributions of LE, LET, and Flat regions on either side of the riblet samples to the frictional drag coefficient.}
    \label{fig:D_F_decompose_bars}
\end{figure}

In addition to the cumulative form of the $D_{f}$ shown in figure \ref{fig:CfCpC3D} (purple bars), the contributions of each side of the sample, as well as the LE, LET, and Flat segments on $D_{f}$ are presented in figure \ref{fig:D_F_decompose_bars}. Firstly, due to the small angle of attack, the contribution of the pressure  side is larger than the suction side for all the samples, including the smooth reference. In addition, for all the riblet samples except for [0.5, 0.5] at ${\rm Re}_L = 18,500$ and $24,200$, and [1.0, -1.0] and [1.5, 0.0] at ${\rm Re}_L = 24,200$, the pressure  sides experience drag increases compared to the smooth reference, and reduction in frictional drag is mostly visible on the suction sides. All members of the ${\cal R} = 1.5$ family experience a reduction in the frictional drag on the suction side. Among the other families, [0.5, -0.5] becomes $D_{f}$-increasing at ${\rm Re}_L = 18,500$ and $24,200$ and [1.0, 0.0] sample is either neutral or $D_{f}$-increasing at ${\rm Re}_L = 12,200$ and $18,500$, while the rest of the cases remain $D_{f}$-decreasing. 

The LE region, captures 14\%, 10\%, and 7.5\% of the length of ${\cal R} = 0.5$, ${\cal R} = 1.0$, and ${\cal R} = 1.5$ families respectively. The LE region experiences the largest levels of local shear stress as is expected from a developing BL. Without any riblets in this area, the drag experienced by the LE regions (dark magenta bars in figure \ref{fig:D_F_decompose_bars}) of all samples of the same constant-${\cal R}$ family experience nearly the same levels of $D_{f}$ and for both sides, the $C_D^{\rm LE}$ are within 0.1\% of the contributions found for the smooth surface ($C_{D,0}^{\rm LE}$). Thus, the LE regions do not contribute to the drag change and this portion of the frictional drag, $C_D^{\rm LE}$, is not available for modification by the riblets. With the difference in the starting point of the riblets, i.e. $x_{\rm LET}$, for the different families, the ${\cal R} = 0.5$ experiences the largest contribution from the $C_D^{\rm LE}$ region, with ${\cal R} = 1.0$ getting a lower contribution, and ${\cal R} = 1.50$ capturing the lowest contribution from the LE region.

Based on $x_{\rm LET}$, 11\%, 15\%, and 17.5\% of the length of the ${\cal R} = 0.5$, ${\cal R} = 1.0$, and ${\cal R} = 1.5$ samples is comprised of the growing riblets in the LET region respectively. Thus, as shown in figure \ref{fig:D_F_decompose_bars}, in reverse order compared with the LE regions, the LET portion of the ${\cal R} = 1.5$ family captures a larger portion of the total $D_{f}$ on either side of the sample, compared to ${\cal R} = 1.0$, and ${\cal R} = 1.0$ having a larger $C_D^{\rm LET}$ than the ${\cal R} = 0.5$ family. With the BL still developing in this region, the magnitudes of the local $\langle \tau_{\rm w} \rangle$ are still large in this segment and while only capturing less than 18\% of the total length, $C_D^{\rm LET}$ captures a considerable portion of the total drag force on either side, especially for the sharper riblets of the ${\cal R} = 1.5$ and ${\cal R} = 1.0$ families. While the changes in the $C_D^{\rm LET}$ is a small portion of the total change the frictional drag experiences, the changes in this region act as a starting point toward drag-reducing or increasing riblets and will be discussed more in section \ref{sec:shear}. The Flat segment capturing 75\% of the length experiences the largest portion of the $C_D^{f}$.

Cumulatively, the drag on the suction sides of all samples and at all Reynolds numbers follow similar trends with respect to $\kappa_2$ as that of the total $D_{f}$ as discussed in section \ref{sec:drag}. On the pressure side however, while ${\cal R} = 1.0$ and $1.5$ follow the same trend as the total $D_{f}$, the frictional drag of the pressure side of ${\cal R} = 0.5$ family is non-monotonic with $\kappa_2$ with the [0.5, 0.0] sample experiencing the largest $D_{f}^{\rm Pressure}$.

\subsection{Local Shear Stress Distribution} \label{sec:shear}

To explain the trend in the friction drag, and its decomposition in terms of the suction/pressure side, as well as the LE, LET, and Flat part of the plate, we focus on the local spanwise-averaged skin friction coefficients distributions, $\langle C_f \rangle(x)$. In general, we see 4 types of $\langle C_f \rangle(x)$ as a function of the local Reynolds number, ${\rm Re}_x = \rho U(x) x/\mu$, where $U(x)$ is the velocity at the edge of the BL and is nearly always larger than $U_{\infty}$. Overall, these $\langle C_f \rangle(x)$ distributions show a more pronounced dependence on the total length of the foil and the location along the plate with respect to that ($x/L$), and the effect of the global Reynolds number (as dictated by $U_{\infty}$) is more visible in the order of magnitude of the values $\langle C_f \rangle(x)$. 
 
\begin{figure}
    \centering
    \includegraphics[width = 1\textwidth]{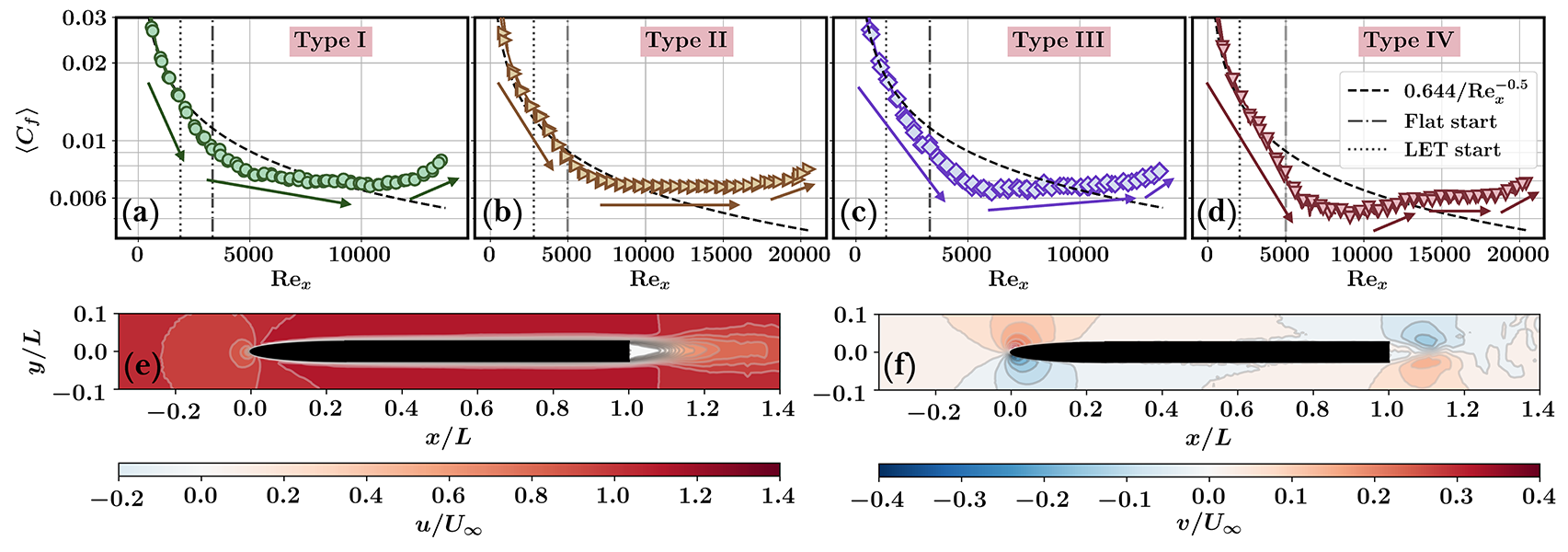}
    \caption{Four types of shear stress distribution observed in flow over riblets and representative examples from the data. (a) Type I, for [0.5, -0.5] sample, at ${\rm Re}_L$ = 12,200 suction side, (b) Type II, for [0.5, 0.0] sample, at ${\rm Re}_L$ = 18,500 pressure side, (c) Type III, for [1.0, 1.0] sample, at ${\rm Re}_L$ = 12,200 suction side, and (d) Type IV, for [1.0, 0.0] sample, at ${\rm Re}_L$ = 18,500 suction side. Colors and markers match the colors and markers used in the upcoming plots. Contour plots of (e) $u$, and (f) $v$, for riblet sample [1.0, 1.0] operated at ${\rm Re}_L = 24,200$.}
    \label{fig:shear_type}
\end{figure}

In the first type (Type I), shown in figure \ref{fig:shear_type}(a), the $\langle C_f \rangle(x)$ first follows a fast decreasing trend (as expected) as a function of the ${\rm Re}_x$, starting in the LE, continuing in the LET and up to early in the Flat region. Afterward, the decreasing trend continues but at a much lower rate than in the first segment (lower than the ${\rm Re}_x^{-1/2}$ of the Blasius BL theory \citep{schlichting2016boundary}), until close to the trailing edge, where the $\langle C_f \rangle(x)$ takes an increasing trend with ${\rm Re}_x$. The smooth sample also follows a Type I trend for all the Reynolds numbers and on both the suction and pressure sides \citep{fu2023doublelightsheet}. In the second type (Type II), shown in figure \ref{fig:shear_type}(b), $\langle C_f \rangle(x)$ starts with a decreasing trend throughout LE, LET, and early Flat region, and then $\langle C_f \rangle(x)$ becomes nearly constant in the Flat region, until close to the trailing edge, $\langle C_f \rangle(x)$ slightly increases. In the third type (Type III), shown in figure \ref{fig:shear_type}(c), $\langle C_f \rangle(x)$ starts in a similar manner as Types I and II, and the difference starts in the second part where after reaching a minimum $\langle C_f \rangle$  in the early portion of the the Flat region, $\langle C_f \rangle (x)$ takes an increasing trend continuing to the trailing edge, with the increase at the trailing edge having a slightly faster rate. In the fourth type (Type IV), shown in figure \ref{fig:shear_type}(d), in the early portion, the same decreasing trend as that of Types I, II, and III is visible, however, in this case, the decrease continues to a minimum, followed by an increase, leading to a region with near constant $\langle C_f \rangle(x)$ or $\langle C_f \rangle (x)$ increasing at a very low rate of change, before getting close to the trailing edge where the $\langle C_f \rangle (x)$ increases at a faster rate. 

\begin{sidewaysfigure}
    \centering
    \vspace{37em}
    \includegraphics[width = 0.95\textheight]{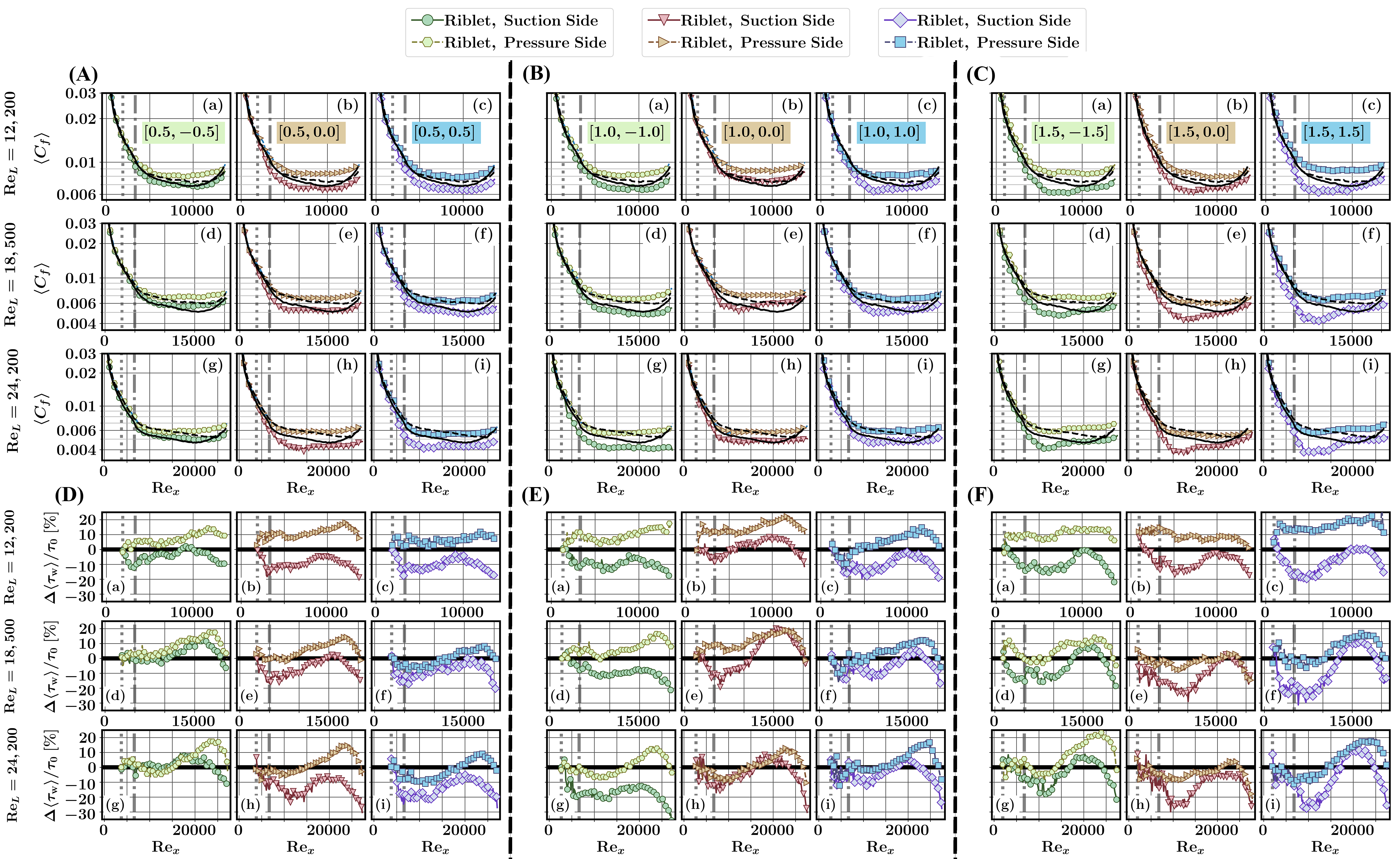}
    \caption{Distribution of $\langle C_f \rangle (x)$ and the percentage of $\Delta \langle \tau_{\rm w} \rangle/\tau_0$ for all the riblet samples of (A) and (D) ${\cal R} = 0.5$, (B) and (E) ${\cal R} = 1.0$, and (C) and (F) ${\cal R} = 1.5$ for all the Reynolds numbers. The $C_{f,0}(x)$ of the reference smooth sample on the suction and pressure sides are shown with solid and dashed black lines respectively. Locations of $x_{\rm LET}$ and $x_{\rm Flat}$ are marked by grey dotted and dash-dotted vertical lines.}
    \label{fig:shear}
\end{sidewaysfigure}

It should be pointed out that in the vicinity of the trailing edge, the $\langle C_f \rangle (x)$ of all samples experiences an increasing trend, different from that usually expected from the BL. This is due to the finite thickness of the sample and how toward the trailing edge, while $u$ maintains a similar behaviour (figure \ref{fig:shear_type}(e)), $v$, (figure \ref{fig:shear_type}(f)) changes sign compared to earlier along the plate, moving toward the body and not away from the surface. In a developing BL, as seen in the earlier portion of the plate, $v$ is always away from the surface (see the color contours of figure \ref{fig:shear_type}(f)), but from $x/L \approx 0.7$, $v$ changes direction toward the plane (change in the color contours) and as a result the velocity profiles are pushed into the surface resulting in more attached BL and thus increase in $\langle C_f \rangle (x)$ for both smooth \citep{fu2023doublelightsheet} and riblet samples. This trend has been reported in second-order models of BL over finite-length plates \mbox{\citep{Dennis85}} as well.

\begin{figure}
    \centering
    \includegraphics[width = \textwidth]{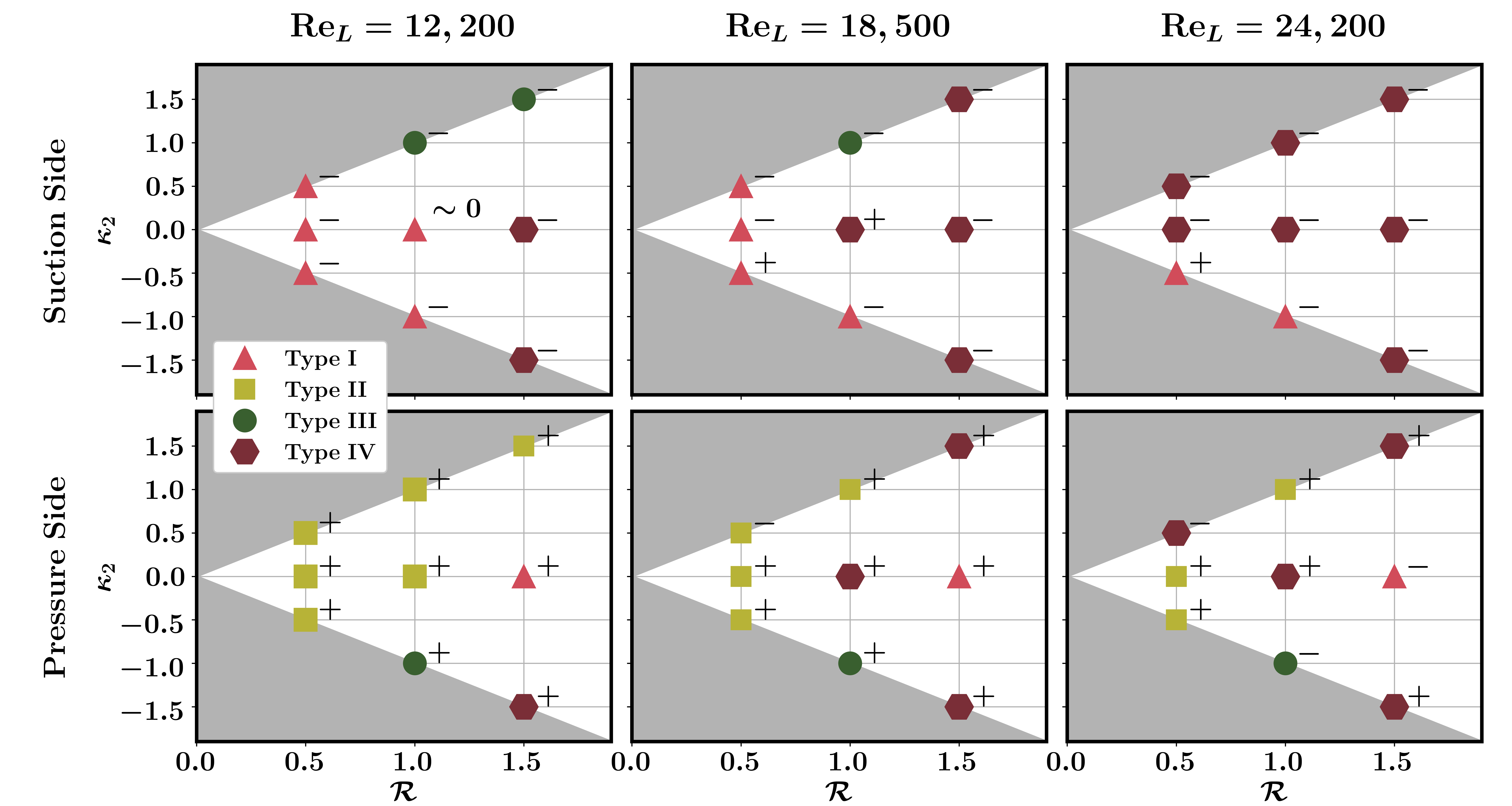}
    \caption{Phase map summarizing the type of distribution of the $\langle C_f \rangle (x)$ as a function of the $\cal R$ and $\kappa_2$ for suction and pressure sides and the global Reynolds numbers. Signs (-/+) on the top right side of the markers indicate the drag-reducing/increasing nature of those sides of the samples (frictional component).}
    \label{fig:phase_map}
\end{figure}

The local distributions of $\langle C_f \rangle (x)$ are presented in figure \ref{fig:shear} for the suction and pressure sides of all the samples and at all global Reynolds numbers, as well as the accompanying distributions of the difference between the spanwise-averaged shear stress of the riblet samples and the smooth reference, $\Delta \langle \tau_{\rm w} \rangle$, normalized locally by that of the smooth reference, $\tau_{_0}$. In addition, a phase map of the type of shear stress distribution as a function of ${\cal R}$ and $\kappa_2$ for all the samples and the global Reynolds numbers is shown in figure \ref{fig:phase_map}. While the $\langle C_f \rangle (x)$ distributions follow only 4 patterns, within each type, we see a variety of the patterns when considering the corresponding $\Delta \langle \tau_{\rm w} \rangle/\tau_0$ distribution as discussed in more details below. For all the cases, prior to the start of the textures, in the LE region, the plates experience $\langle C_f \rangle (x)$ nearly identical to that of the smooth reference plate ($C_{f,0}(x)$) as expected. Hence, we record similar values of $C_D^{\rm LE}$ for all the members of each family and the smooth surface as was shown in section \ref{sec:friction}. As the riblets start and grow in the LET region, the $\langle C_f \rangle (x)$ starts to deviate from that of the smooth reference. As discussed earlier in section \ref{sec:sample}, the design of the leading edge mimicking the nose of a shark \citep{lauder2016structure}, with a smooth LE region and gradual growth of the riblets in the LET region, allows the $\langle C_f \rangle (x)$ to incrementally develop with the growth of the textures, and thus  avoiding the large levels of shear stress as seen in the leading edge of fully-covered plates \citep{raayai2017drag}. Due to this development, in the LET we see differences between the $D_{f}^{\rm LET}$ contributions recorded for the riblet samples and the smooth; on the suction  sides, the LET regions of the majority of samples capture some level of local shear reduction, while on the pressure  side, the riblets capture both shear reductions and increases.

The Type I shear stress distribution (figures \ref{fig:shear} and \ref{fig:phase_map}) is mainly seen among the members of the shallowest family of ${\cal R} = 0.5$, as shown in figure \ref{fig:shear}(Aa-g) and \ref{fig:shear}(Da-g), dominating the suction sides of ${\rm Re}_L = 12,200$ and ${\rm Re}_L = 18,500$ cases for all the members, independent of $\kappa_2$, and the [0.5, -0.5] sample at ${\rm Re}_L = 24,200$. For sharper riblets, this type only is seen to persist on the suction side of [1.0, 0.0] sample at ${\rm Re}_L = 12,200$ (figure \ref{fig:shear}(Bb) and \ref{fig:shear}(Eb)), and  at all tested Reynolds number on the suction side of [1.0, -1.0] sample (figure \ref{fig:shear}(Ba,d,g), and figure \ref{fig:shear}(Ea,d,g)) and pressure side of the [1.5, 0.0] sample (figure \ref{fig:shear}(Cb,e,h), and figure  \ref{fig:shear}(Fb,e,h)). Among one group of the cases, on the suction sides of all the ${\cal R} = 0.5$ samples at ${\rm Re}_L = 12,200$, and [0.5, 0.0] and [0.5, 0.5] at ${\rm Re}_L = 18,500$, as well as the [1.0, -1.0] sample at all the tested Reynolds numbers, in the LET region, $\langle C_f \rangle$ grows to lower values than the $C_{{f},0}$ and after this initial reducing trend in the $\Delta \langle \tau_{\rm w} \rangle/\tau_0$ in the LET region, samples experience a lower rate of decrease in the $\langle C_f \rangle (x)$ as a function of the ${\rm Re}_x$ compared with the smooth reference and thus they see increasing trends in $\Delta \langle \tau_{\rm w} \rangle/\tau_0$, at times reaching close to the $C_{f,0}(x)$, until about $x/L \approx 0.8$ where the trailing edge effect sets in and the increasing trend in the $C_{f,0}(x)$ of the smooth reference takes over the increasing trend of $\langle C_f \rangle$ of the riblet samples. This results in both $D_{\rm Suction}^{\rm LET}$ and $D_{\rm Suction}^{\rm Flat}$ components of these cases to experience reductions as shown in figure \ref{fig:D_F_decompose_bars}, even with the increasing trend in the $\Delta \langle \tau_{\rm w} \rangle/\tau_0$ in the Flat region, and ultimately resulting in overall drag reduction on the suction sides of these samples. In a second group, as we see on the suction side of [0.5, -0.5] at ${\rm Re}_L = 18,500$ and ${\rm Re}_L = 24,200$, and pressure side of the [1.5, 0.0] sample at ${\rm Re}_L=12,200$, $\langle C_f \rangle$ grows to similar or larger values than the $C_{{f},0}$ in the LET, and after the initial $\Delta \langle \tau_{\rm w} \rangle/\tau_0 \geq 0$ region, with a lower rate of decrease in $\langle C_f \rangle (x)$ compared to the smooth reference along the length, $\Delta \langle \tau_{\rm w} \rangle/\tau_0$ takes an increasing trend in the Flat region prior to decreasing in the trailing edge. Overall, along most of the length, $\Delta \langle \tau_{\rm w} \rangle/\tau_0 \geq 0$, hence these samples stay $D_f$-increasing as shown in figure \ref{fig:D_F_decompose_bars}. For the rest of the cases, namely pressure side of [1.5, 0.0] at ${\rm Re}_L = 18,500$ and ${\rm Re}_L = 24,200$ and suction side of the [1.0, 0.0] at ${\rm Re}_L = 18,500$, in the LET, $\langle C_f \rangle (x)$ starts at lower than the smooth reference and later in the Flat region, $\langle C_f \rangle$ crosses over the $C_{f,0}(x)$, turning shear-increasing, until later in the trailing edge area that some move to slightly shear-reducing. As a result of this, suction side of the [1.0, 0.0] at ${\rm Re}_L = 18,500$ stays nearly neutral in terms of the $D_f$ changes, and pressure side of [1.5, 0.0] at ${\rm Re}_L = 18,500$ is $D_f$-increasing, while at ${\rm Re}_L = 24,200$, the extent of the $\Delta \langle \tau_{\rm w} \rangle <0$ is large enough that the pressure side of [1.5, 0.0] becomes $D_f$-reducing (one of the few cases where pressure side is drag reducing).

Type II shear stress distributions are only seen on the pressure side and they only experience either drag increases or no change compared with the smooth counterpart except for one case of pressure side of [0.5, 0.5] at ${\rm Re}_L = 18,500$ which records a 1.2\% reduction. This type (see figure \ref{fig:phase_map}) is seen on the pressure side of the ${\cal R} = 0.5$ family at all Reynolds numbers, except for [0.5, 0.5] sample at ${\rm Re}_L = 24,200$ (figure \ref{fig:shear}(Aa-Ah) and \ref{fig:shear}(Da-Dh)), as well as the pressure sides of [1.0, 0.0] at ${\rm Re} = 12,200$ (figure \ref{fig:shear}(Bb) and \ref{fig:shear}(Eb)), [1.0, 1.0] at all tested Reynolds numbers (figure \ref{fig:shear}(Bc,f,i) and \ref{fig:shear}(Ec,f,i)), and [1.5, 1.5] at ${\rm Re} = 12,200$ (figure \ref{fig:shear}(Cc) and \ref{fig:shear}(Fc)). In one group of the cases, such as the pressure sides of [0.5, -0.5] and [0.5, 0.0] at ${\rm Re}_L = 12,200$ and $18,500$, [0.5, 0.5] at ${\rm Re}_L = 12,200$, as well as [1.0, 0.0] and [1.5, 1.5] at ${\rm Re}_L = 12,200$, in the LET region, the $\langle C_f \rangle$ grows to larger than or near equal values to the smooth $C_{f,0}(x)$ and as $\langle C_f \rangle (x)$ becomes constant, $\Delta \langle \tau_{\rm w} \rangle/\tau_0$ keeps increasing, resulting in $D_{\rm Pressure}^{\rm LET}$ and $D_{\rm Pressure}^{\rm Flat}$ larger than or near equal to the smooth ones. In the second group of cases, [0.5, -0.5] and [0.5, 0.0] at ${\rm Re}_L = 24,200$, while in the LET, the $\langle C_f \rangle \geq C_{{f},0}$, toward the end of the LET and early Flat, $\langle C_f \rangle (x)$ moves to lower than the smooth reference with a small region experiencing $\Delta \langle \tau_{\rm w} \rangle/\tau_0<0$. However, as $\langle C_f \rangle (x)$ becomes constant, it crosses over the smooth reference, ultimately leading to both $D_{\rm Pressure}^{\rm LET}$ and $D_{\rm Pressure}^{\rm Flat}$ larger than or near equal to the smooth ones. In the last group, on the pressure side of the [0.5, 0.5] at ${\rm Re}_L = 18,500$, [0.5, 0.0] at ${\rm Re}_L = 24,200$, and [1.0, 1.0] at all Reynolds numbers, initially, $\langle C_f \rangle$ grows to lower than $C_{f,0}(x)$ of the smooth reference but as the $\langle C_f \rangle (x)$ becomes constant, it crosses over the smooth becoming shear-increasing and while $D_{\rm Pressure}^{\rm LET}$ is smaller than the smooth reference, the shear-increasing portion of the Flat region, except for the [0.5, 0.5] case at ${\rm Re}_L = 18,500$, results in $D_{\rm Pressure}^{\rm Flat}$ going to larger than the smooth one, becoming cumulatively $D_{f}$-increasing.

Type III distribution is the least common trend, mainly observed on the suction sides of concave textures [1.0, 1.0] at ${\rm Re} = 12,200$ and $18,500$ (figure \ref{fig:shear}(Bc,f), and \ref{fig:shear}(Ec,f)), and [1.5, 1.5] at ${\rm Re} = 12,200$ (figure \ref{fig:shear}(Cc), and \ref{fig:shear}(Fc)), and on the pressure side of the convex sample [1.0, -1.0] for all the tested Reynolds numbers (figure \ref{fig:shear}(Ba,d,g), and \ref{fig:shear}(Ea,d,g)). On the suction sides, all these cases are $D_{f}$-reducing, while on the pressure side, all, besides one, are $D_{f}$-increasing. In the Type III cases on suction sides, from the LET region, the samples start in a shear-reducing pattern which continues to the point where $\langle C_f \rangle$ reaches its minimum and afterward as $\langle C_f \rangle (x)$ starts to increase, $\Delta \langle \tau_{\rm w} \rangle$ also takes an increasing trend either getting marginally close to the smooth reference or crossing over the $C_{{f},0}$ for a slight bit, before moving to shear-reducing at the trailing edge. With a considerable extent of the LET and Flat region staying in shear-reducing conditions, the $D_{\rm Suction}^{\rm LET}$ and $D_{\rm Suction}^{\rm Flat}$ stay lower than the smooth reference. On the pressure side, [1.0, -1.0] sample at ${\rm Re}_L = 12,200$ and $18,500$, stay consistently shear-increasing, with the increasing trend of the $\langle C_f \rangle (x)$ along the length past the minimum $\langle C_f \rangle$ resulting in a consistent increase in the $\Delta \langle \tau_{\rm w} \rangle$ in the Flat region, thus keeping both these cases as $D_{f}$-increasing. This is while [1.0, -1.0] at ${\rm Re}_L = 24,200$ start in a shear-reducing state in the LET, but in the middle of the Flat region, $\langle C_f \rangle (x)$ crosses over the $C_{f,0}(x)$ of the smooth, becoming shear-increasing in the latter part. However, the extent of the shear-reducing region is able to maintain a $D_{f}$-reducing behaviour (one of the few observed on the pressure side).

Type IV distribution (figures \ref{fig:shear} and \ref{fig:phase_map}), is mostly dominant on the suction sides of nearly all samples at ${\rm Re}_L = 24,200$ except for [0.5, -0.5] and [1.0, -1.0] (figure \ref{fig:shear}(Ah-i), \ref{fig:shear}(Bh-i), \ref{fig:shear}(Cg-i), and \ref{fig:shear}(Dh-i), \ref{fig:shear}(Eh-i), and \ref{fig:shear}(Fg-i)). Independent of the Reynolds numbers, Type IV is dominant on the suction sides of the members of the ${\cal R} = 1.5$ family except the [1.5, 1.5] sample at ${\rm Re}_L = 12,200$ and $18,500$ (figure \ref{fig:shear} (Ca,b,d,e,f), and \ref{fig:shear}(Fa,b,d,e,f)). Besides these two larger groups, suction side of [1.0, 0.0] at ${\rm Re}_L = 18,500$ (figure \ref{fig:shear}(Be) and \ref{fig:shear}(Ee)) has a Type IV distribution. On the pressure side, [0.5, 0.5] at ${\rm Re}_L = 24,200$ (figure \ref{fig:shear}(Ai) and \ref{fig:shear}(Di)), [1.0, 0.0] and [1.5, 1.5] at ${\rm Re}_L = 18,500$ and $24,200$ (figure \ref{fig:shear}(Be,h), \ref{fig:shear}(Ee,h), \ref{fig:shear}(Cf,i), and \ref{fig:shear}(Ff,i)), and [1.5, -1.5] at all tested Reynolds numbers (figure \ref{fig:shear}(Ca,d,g) and \ref{fig:shear}(Fa,d,g)) experience a Type IV distribution. Overall, the samples that experience any of the Types I-III distribution at the lowest Reynolds number of $12,200$, either sustain the same distribution type at the larger global Reynolds numbers ($18,500$, and $24,200$) or ultimately transition to a Type IV as the global Reynolds number is increased. Samples [-1.5, 1.5] on both sides and [1.5, 0.0] on the suction side capture a Type IV distribution for all the tested Reynolds numbers.  

Among all Type IV distributions on the suction sides, we see the $\langle C_f \rangle (x)$ in the LET region growing to levels lower than the $C_{f,0}(x)$ of the smooth surface, offering shear reduction throughout the LET and resulting in $D_{\rm Front}^{\rm LET}$ to be lower than that of the smooth. Then in the early portion of the Flat region, $\langle C_f \rangle (x)$ keeps its fast rate of  decrease along the length of the plate (while the rate of decrease of $C_{{f},0}$ of the smooth has started to decline), thus increasing the distance between the $\langle C_f \rangle (x)$ and $C_{{f},0}(x)$, recording the largest local shear reduction as much as 20\% for suction sides of [0.5, 0.0] and [0.5, 0.5] samples, and 30\% for suction sides of the [1.5, 0.0] and [1.5, 1.5] cases at ${\rm Re}_L = 24,200$, and 12\% for suction side of the [1.0, 0.0] sample at ${\rm Re}_L = 18,500$, and about 15\% for suction sides of the [1.0, 0.0] and [1.0, 1.0] samples at ${\rm Re}_L = 24,200$. Local shear stress reductions ranging 20-50\% has been previously reported by \citet{furuya1977turbulent}, \citet{hooshmandi983}, and \citet{gallagher1984turbulent}. After reaching its global minimum, the $\langle C_f \rangle (x)$ then takes on an increasing trend which is deterministic of how much $D_{f}$ reduction can be possible for these samples. In cases such as suction side of the [1.5, 0.0] sample at all Reynolds numbers, along the increasing and then constant path of the $\langle C_f \rangle (x)$ in the $x$ direction, the $\langle C_f \rangle (x)$ comes marginally close to the $C_{{f},0}(x)$, but always staying at $\langle C_f \rangle(x) <C_{f,0}(x)$, never crossing over, thus, even though non-monotonic and at times not optimal, it maintains a consistent $\Delta \langle \tau_{\rm w} \rangle \leq 0$ all along the length. These cases stay drag-reducing, in both $D_{\rm Front}^{\rm LET}$ and $D_{\rm Front}^{\rm Flat}$ components and cumulatively along the length. In other cases, such as the suction side of the [1.5, 1.5] and [1.5, -1.5] at ${\rm Re}_L = 18,500$ and $24,200$, along the increasing and then constant path of the $\langle C_f \rangle (x)$, the $\langle C_f \rangle (x)$ is able to cross over the the $C_{{f},0}$, with portions of the sample experiencing $\langle C_f \rangle(x) > C_{{f},0}(x)$. As a result, along the length, the sample starts at $\Delta \langle \tau_{\rm w} \rangle \leq 0$ in the LET and first half of the Flat region, then crossing over to $\Delta \langle \tau_{\rm w} \rangle > 0$ in parts of the second half of the Flat region, before a small region in the trailing edge with $\Delta \langle \tau_{\rm w} \rangle \leq 0$. For these cases, with the positive and negative $\Delta \langle \tau_{\rm w} \rangle$ crossing each other out, it is not easy to determine the drag-reducing/increasing nature of the samples without calculation of the integrals of equation \eqref{contributions}. Thus, all the Type IV distributions on the suction side lead to drag reductions besides the one case of the [1.0, 0.0] sample at ${\rm Re} = 18,500$ which experiences a $\langle C_f \rangle (x)$ crossing over the smooth reference and a substantial shear-increase as shown in figure \ref{fig:shear}(Ee).

On the pressure side, the [0.5, 0.5], [1.0, 0.0], and [1.5, 1.5] samples at ${\rm Re}_L = 24,200$ follow a very similar trend as that of suction sides described above, however, both experience a cross-over to $\Delta \langle \tau_{\rm w} \rangle>0$, where only [0.5, 0.5] is able to maintain a $D_{f}$-reducing behaviour and the shear-increasing region of [1.0, 0.0], and [1.5, 1.5], turn both cases to be $D_{f}$-increasing (see figure \ref{fig:D_F_decompose_bars}). Pressure side of the [0.5, 0.5] at ${\rm Re}_L = 24,200$ is the only case of a Type IV distribution that is drag-reducing on the pressure side. As for the rest, in one group of the Type IV cases on the pressure side such as the cases of [1.5, 1.5] and [1.5, -1.5] at ${\rm Re}_L = 18,500$, initially, we have the $\langle C_f \rangle (x)$ grow to larger than the $C_{f,0}(x)$ of the smooth case, but close to the end of the LET and start of the Flat region, $\langle C_f \rangle (x)$ crosses over to $\Delta \langle \tau_{\rm w} \rangle\leq0$, before turning shear increasing toward the middle of the Flat region. Another group of the Type IV cases, [1.0, 0.0] at ${\rm Re}_L = 24,200$ and [1.5, -1.5] at ${\rm Re}_L = 12,200$ stay in their entirety on the shear-increasing side and thus are fully $D_{f}$-increasing in both LET and Flat regions as well as the entirety of the pressure sides of the plates.

\subsection{Distribution of the Fitting Parameters} \label{fittingParameterDist}
The distribution of the $\langle C_f \rangle (x)$, in presence of the riblets, as a function of the ${\rm Re}_x$ (shown in section \ref{sec:shear}), can be explained further by looking at the distributions of the two fitting parameters $m$ and $n_0$ for all the samples at all the tested global Reynolds numbers. The distribution of the two parameters, show an intertwined dependence on each other and the cross-sectional area available to the flow inside the grooves. With the $\langle C_f \rangle (x)$, already capturing the effect of the increase in the wetted surface area of the riblets compared with a smooth surface, as well as the changes in the shear stress distribution inside the grooves, the distributions of $m$ and $n_0$ are also cumulative views of the flow dynamics and the geometric variations imposed by the presence of the riblets. The high frequency variations in the data presented in this and subsequent sections has been filtered with a Savitzky-Golay filter \citep{savitzky1964smoothing, savitzky1989historic}.

\subsubsection{Fitting parameter $m$} \label{sec:m}
\begin{sidewaysfigure}
    \centering
    \vspace{37em}
    \includegraphics[width = 1\textheight]{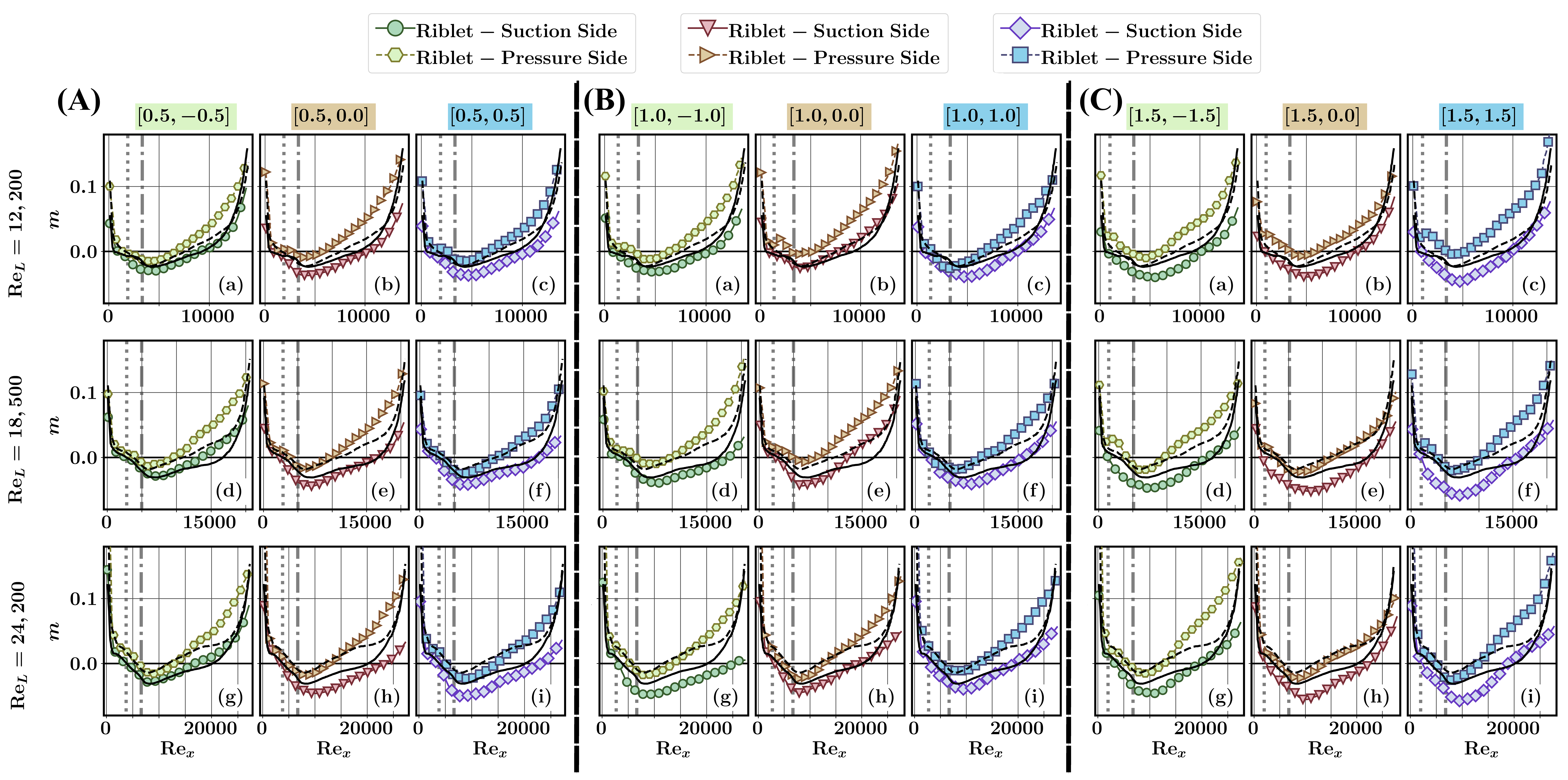}
    \caption{Distribution of $m$ for all the riblet samples of (A) ${\cal R} = 0.5$, (B) ${\cal R} = 1.0$, and (C) ${\cal R} = 1.5$ and the reference smooth samples, on the suction  and pressure  sides for all the tested Reynolds numbers. The $m$ of the reference smooth sample on the suction and pressure sides are shown with solid and dashed black lines respectively.}
    \label{fig:m}
\end{sidewaysfigure}

The fitting parameter $m$ (presented in figure \ref{fig:m} for all cases) plays a significant role in capturing the local effect of multiple physical phenomena in the flow field; the pressure gradient, the effect of both the limited (i.e. short) length of the plate and the presence of the riblets on the viscous diffusion, as well the effect of the spanwise-averaging operation on the non-linear advection terms. For all the cases, including the smooth samples, and on both sides, the distribution of $m$ is comprised of three distinct regions; first a $m>0$ region in the leading edge area, then the $m<0$ area in the middle including the location where $m$ reaches a global minimum, and lastly  the $m>0$ for the rest of the length of the plate toward the trailing edge. As is seen in figure \ref{fig:m}, $m$ is larger on the pressure side than the suction side and as a result, the extent of the $m<0$ region varies quite a bit between the sides and samples, and can take small or large portion of the length of the plate. In general, $m$ takes a decreasing trend in the early portion of the plate prior to the global minimum and then an increasing trend in the second half toward the trailing edge. The initial decreasing trend is not strictly monotonic for many of the cases on the pressure side such as those of [0.5, 0.0] (figure \ref{fig:m}(Ab)), [0.5, 0.5](figure \ref{fig:m}(Ac)), [1.0, -1.0] (figure \ref{fig:m}(Ba)), and [1.5, 0.0] (figure \ref{fig:m}(Cb)) at ${\rm Re}_L = 12,200$, [1.0, 0.0] (figure \ref{fig:m}(Bb,e)) and [1.5, 1.5] (figure \ref{fig:m}(Cc,f)) at ${\rm Re}_L = 12,200$ and $18,500$, and [1.5, -1.5] (figure \ref{fig:m}(Ca,d,g)) at all the Reynolds numbers where in the LE/LET regions there is a local increasing-decreasing trend that sets in as the flow sees the riblets for the first time.

In a FS type of BL equations, at a constant Reynolds number, increasing $m$ monotonically increases the wall shear stress, while at a constant $m$, increasing the Reynolds number monotonically reduces the wall shear stress. Thus, for all the cases, while prior to the minimum $m$, the increase in the ${\rm Re}_x$ and decrease in $m$ work together in reducing the $\langle C_f \rangle (x)$ along the length, the evolution of the $m$ past the minimum is the determining factor for the type of the shear stress distribution each sample takes. After the minimum $m$, along the length of the sample, the increase in the local Reynolds number results in the wall shear stress decreasing, however, the increase in the $m$ disrupts this trend and depending on the rate of increase of $m$, the trend in the $\langle C_f \rangle (x)$ is set.

In the Type I distribution, the effect of rate of increase in $m$ on the $\langle C_f \rangle (x)$, is not able to overcome the effect of the ${\rm Re}_x$ in decreasing the $\langle C_f \rangle (x)$ along the length and thus overall, we see a decreasing trend in the $\langle C_f \rangle (x)$ along the length prior to the vicinity of the trailing edge albeit at a lower rate than the region prior to the minimum $m$ and lower than the ${\rm Re}_x^{-1/2}$ of the BL theory. However, as it is seen in figure \ref{fig:m}(Aa-g) (suction sides) for ${\cal R} = 0.5$ family, figure \ref{fig:m}(Ba,b,d,g) (suction sides) for ${\cal R} = 1.0$ family, figure \ref{fig:m}(Cb,e,h) (pressure sides) for ${\cal R} = 1.5$ family, the rate of increase in $m$ is faster for the riblets than the smooth reference, thus reducing the rate of decrease of $\langle C_f \rangle (x)$ along the length. After the minimum $m$, the $\Delta \langle \tau_{\rm w} \rangle/\tau_0$ takes an increasing trend until the vicinity of the trailing edge, at around $x/L \approx 0.8$, where the rate of increase in $m$ increases more, leading to $\langle C_f \rangle (x)$ of both the riblet and smooth samples increasing (discussed earlier). In the Type II distribution, the increase in $m$ nearly balances that of the increase in the ${\rm Re}_x$, resulting in nearly constant $\langle C_f \rangle (x)$ prior to the trailing edge effect, as seen in pressure sides of most of the ${\cal R} = 0.5$ family (figure \ref{fig:m}(Aa-h)), and a few cases of the ${\cal R} = 1.0$ (figure \ref{fig:m}(Bb,c,f,i)) and [1.5, 1.5] (figure \ref{fig:m}(Cc)). The difference between the rate of increase of $m$ in Types I and II distributions is clearly visible for the suction and pressure sides of the ${\cal R} = 0.5$ family at ${\rm Re}_L = 12, 200$ and $18,500$ (figure \ref{fig:m}(Aa-f)).  

In Type III and IV distributions, the effect of the increase in $m$ fully overcomes and surpasses that of the local Reynolds number and results in an increasing (and/or near constant) trend in the $\langle C_f \rangle (x)$ along the length. Among these two types, the location of the minimum $m$ is very close to the location of the minimum $\langle C_{f} \rangle(x)$. Specifically for those samples that experience these two types on the suction sides, the initial decreasing trend in $m$ along the length of the sample has a faster rate than that of the smooth sample, resulting in a larger separation between the minimum $m$ of the riblet samples and the smooth reference (for example see suction sides of the members of the ${\cal R} = 1.5$ family, figure \ref{fig:m}(Ca-i)), and resulting in large levels of shear reduction prior to the minimum $\langle C_f \rangle (x)$. Thus, even with the counteracting effect of the increase in ${\rm Re}_x$ and $m$ after the minimum $m$, these samples can maintain a $D_{f}$-reducing trend on the suction sides. Only on the suction side of the [1.0, 0.0], the rate of increase of $m$ is substantially high enough to result in a  $D_{f}$-increasing behaviour.

Overall, the pressure  sides of the samples capture larger values of $m$ than the suction  sides. With the non-linear impact of the rate of change of $m$ on the rate of change in the $C_f$ (keeping all other parameters constant), we see that while the rate of change of $m$ along the length is quite similar between the suction and pressure sides, depending on the sign and magnitude of the $m$, this translates into different rates of change in $\langle C_f \rangle (x)$ between the two sides of the same sample. Mathematically one can see that as $m$ is increased the variations in the $C_f$ slowly decreases; for example, increasing $m$ from -0.02 to -0.01 results in a $\Delta C_f/ C_f(m=0) = 0.066$ (normalized by the $C_f$ of $m=0$ FS solution), while increasing $m$ from 0.01 to 0.02 results in $\Delta C_f/ C_f(m=0) = 0.056$. Thus, comparing the two sides of the samples, with $m_{\rm pressure} \geq m_{\rm suction}$, on the pressure sides the rate of increase in $\langle C_f \rangle (x)$ due to the increase in $m$ is weaker than on the suction side and therefore, with the increase in the ${\rm Re}_x$, the changes in $\langle C_f \rangle (x)$ on the suction side (as well as the $\Delta \langle \tau_{\rm w} \rangle/\tau_0$) show larger variations along the length of the sample compared to those on the pressure sides. This leads to seeing more Type I/IV distributions on the suction side (24/27 riblet cases) and more Type II distributions on the pressure side (13/27) and a few Type IV cases with smaller variations compared to those of the suction sides (8/27)(see figure \ref{fig:phase_map}).

To explore the effect of $m$ further, we perform the spanwise-averaging operation on the Navier-Stokes equation in the $x$ direction for flow past a generic riblet surface and organize and write it in a form resembling that of the BL equation (derivation in appendix \ref{appB} for the Flat region  and appendix \ref{appC} for the curved LE and LET regions)

\begin{equation}
    \rho \left( \langle u \rangle \dfrac{\partial \langle u \rangle}{\partial x} + \langle v \rangle \dfrac{\partial \langle u \rangle}{\partial y}  \right) = - \dfrac{\partial \langle P^{*} \rangle}{\partial x} + \mu  \dfrac{\partial^2 \langle u \rangle}{\partial y^2} \label{NSE_average2}
\end{equation}

\noindent where in the Flat region ($x>x_{\rm Flat}$)

\begin{equation}
    -\dfrac{\partial \langle P^{*} \rangle}{\partial x} = - \underbrace{\dfrac{\partial \langle p \rangle}{\partial x}}_{\raisebox{.5pt}{\textcircled{\raisebox{-.9pt} {1}}}} + \underbrace{\mu \dfrac{\partial^2 \langle u \rangle}{\partial x^2 }  + {\mathcal{Z}}}_{\raisebox{.5pt}{\textcircled{\raisebox{-.9pt} {2}}}}  + \underbrace{\rho \left( \langle u\rangle  \dfrac{\partial \langle u \rangle}{\partial x} + \langle v \rangle\dfrac{\partial \langle u \rangle}{\partial y} - \dfrac{\partial \langle u u \rangle}{\partial x} - \dfrac{\partial \langle u v \rangle}{\partial y}   \right)} _{ \raisebox{.5pt}{\textcircled{\raisebox{-.9pt} {3}}}}\label{updated_p1}
\end{equation}

\begin{equation}
    {\mathcal{Z}} = \left\{
    \begin{alignedat}{3} 
        & - \dfrac{2\mu}{\lambda} \dfrac{\partial u}{\partial z} \bigg \vert_{z=z_{\rm w1}^+}  \ \ \ \ \ & y_{\rm trough} \leq y \leq y_{\rm peak} = h/2 \\
        &0 \ \ \ \ \ &\vert y \vert  > y_{\rm peak}.
    \end{alignedat}
    \right. \label{Z}
\end{equation}

\noindent and in the riblet-covered leading edge, LET, ($x_{\rm LET} < x \leq x_{\rm Flat}$) 

\begin{equation}
    \begin{split}
    - \dfrac{\langle P^* \rangle}{\partial s} =  & - \underbrace{\dfrac{\partial \langle p \rangle}{\partial s}}_{\raisebox{.5pt}{\textcircled{\raisebox{-.9pt} {1}}}}  + \underbrace{\mu \left( \dfrac{\partial^2 \langle u_s \rangle}{\partial s^2} \right) + {\mathcal{Z}} }_{\raisebox{.5pt}{\textcircled{\raisebox{-.9pt} {2}}}} \\ & \underbrace{\rho \left( \langle u_s \rangle \dfrac{\partial \langle u_s \rangle}{\partial s} + \langle v_n \rangle \dfrac{\partial \langle u_s \rangle}{\partial n} - \dfrac{\partial \langle u_s u_s \rangle}{\partial s} -  \dfrac{\partial \langle u_s v_n \rangle}{\partial n}  \right)}_{\raisebox{.5pt}{\textcircled{\raisebox{-.9pt} {3}}}} +  \underbrace{{\mathcal{K}_2}}_{\raisebox{.5pt}{\textcircled{\raisebox{-.9pt} {4}}}}
    \end{split}
\end{equation} 

\noindent replacing $y$ with $n$ and $u$ with $u_s$ in the definition of the ${\mathcal{Z}}$ in equation \ref{Z}, where $u_s$ and $v_n$ are the tangential and normal velocities along the $s$, and $n$, directions with respect to the wall of the curved leading edge (see figure \ref{fig:riblets}), and ${\mathcal{K}_2}$ is the contribution from the curvature of the leading edge as shown in appendix \ref{appC}. In this format, all the excess terms compared with the BL equation are captured via an \emph{equivalent} pressure gradient term, defined as $\partial \langle P^* \rangle/ \partial x$ (or $\partial \langle P^* \rangle/ \partial s$), so that equation \eqref{NSE_average2} matches the form of the FS family of BL equations. Here, locally, by fitting the collected data to the FS equations, we capture the $\partial \langle P^* \rangle /\partial x$ term via the $m$ parameter which is then written in the form of $-{\partial \langle P^* \rangle}/{\partial x} = \rho U(x)^2 {m}/{x}$, where the distribution of this term follows the same trend as the $m$ and and the $1/x$ acts as a scaling factor which decreases along the length. Note that as mentioned in section \ref{shearmethod}, for simplicity, in the LET and LET, we perform the fitting with $x$ instead of $s$ and let $m$ also capture the effect of this difference and thus use $-{\partial \langle P^* \rangle}/{\partial s} = \rho U(x)^2 {m}/{x}$, in the curved leading edge area. Thus, the $m$ parameter, represents the cumulative effect of \raisebox{.5pt}{\textcircled{\raisebox{-.9pt} {1}}} the pressure gradient, \raisebox{.5pt}{\textcircled{\raisebox{-.9pt} {2}}} the non-negligible viscous diffusion term in the x direction and the viscous diffusion term in the z direction due to the 3D effect inside the grooves, as well as \raisebox{.5pt}{\textcircled{\raisebox{-.9pt} {3}}} the residues from the non-linear terms left from the averaging operation. In the leading edge area $m$ also captures the effect of the curvature both through the ${\mathcal{K}_2}$ term (see appendix \ref{appC}) and the use of $x$ in the fitting process. In the LE, equation \eqref{NSE_average2} turns back into (see appendix \ref{appC})

\begin{equation}
    \rho \left( u_s  \dfrac{\partial u_s }{\partial s} + v_n  \dfrac{\partial u_s }{\partial n}  \right) = - \dfrac{\partial P^{*} }{\partial s} + \mu  \dfrac{\partial^2 u_s }{\partial n^2}  \ \ \ \ \ ; \ \ \ -\dfrac{\partial P^{*} }{\partial s} = - \underbrace{\dfrac{ \partial p }{\partial s}}_{\raisebox{.5pt}{\textcircled{\raisebox{-.9pt} {1}}}} + \underbrace{\mu \dfrac{\partial^2 u_s }{\partial s^2 }}_{\raisebox{.5pt}{\textcircled{\raisebox{-.9pt} {2}}}}   + \underbrace{ {\mathcal{K}_1}   }_{\raisebox{.5pt}{\textcircled{\raisebox{-.9pt} {4}}}} \label{NSE_average_curve}  
\end{equation}

\begin{sidewaysfigure}
    \centering
    \vspace{37em}
    \includegraphics[width = 1\textheight]{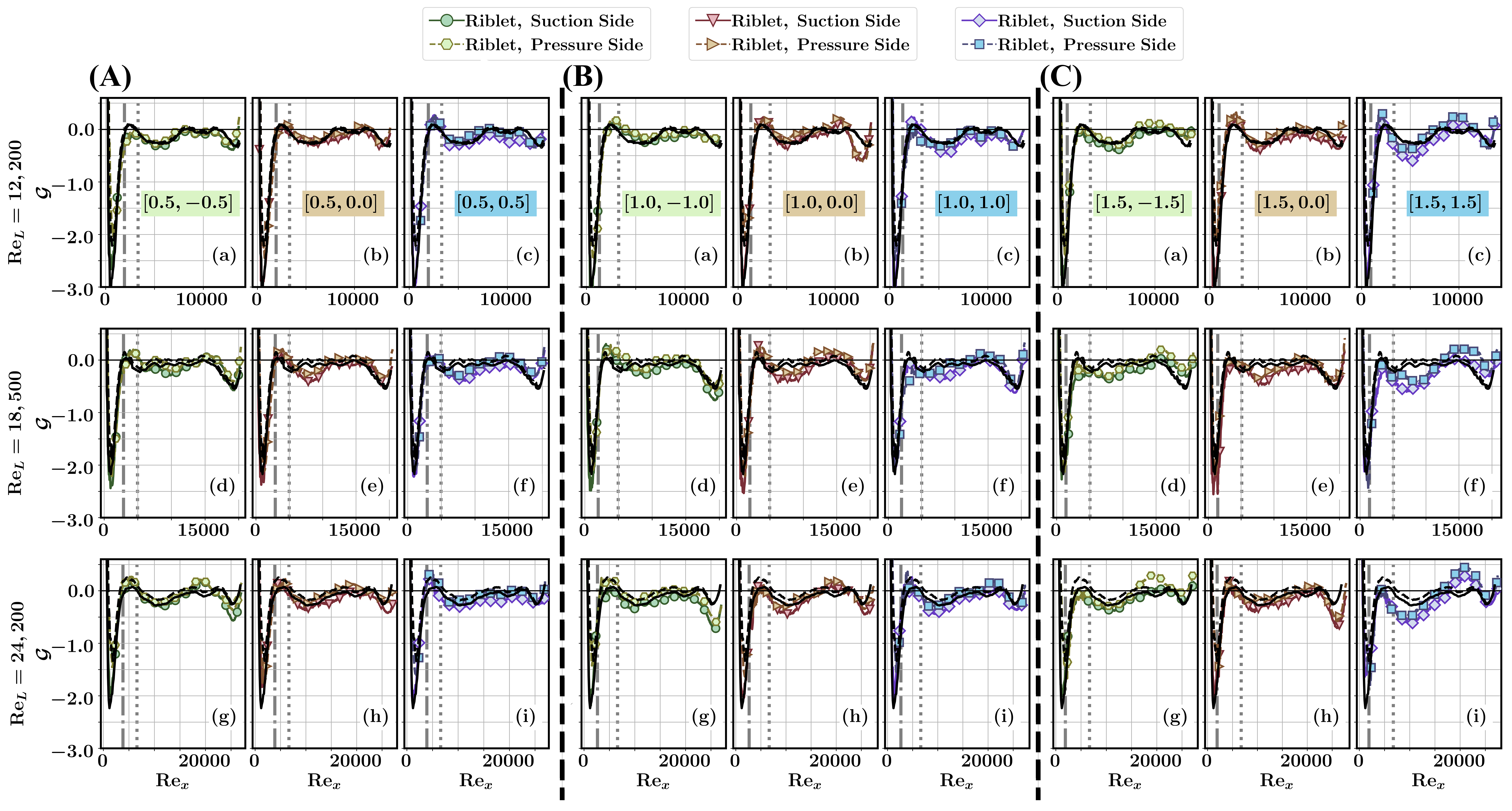}
    \caption{Distribution of ${\cal G}$ or difference between the $\partial \langle P^* \rangle/\partial x$ and $\partial \langle p \rangle/\partial x$ terms in dimensionless form, for all the riblet samples of (A) ${\cal R} = 0.5$, (B) ${\cal R} = 1.0$, and (C) ${\cal R} = 1.5$ on the suction  and pressure  sides of the riblet samples. The results for the smooth reference for all the tested Reynolds numbers are shown with solid and dashed black lines for the suction and pressure sides respectively.}
    \label{fig:p_grad_diff}
\end{sidewaysfigure}

\noindent as expected missing the ${\raisebox{.5pt}{\textcircled{\raisebox{-.9pt} {3}}}}$ term compared with the riblet regions with $-{\partial P^*}/{\partial s} = \rho U(x)^2 ({m}/{x})$. Separate from the curve fitting operation, we use the PIV data and find the pressure gradient following section \ref{sec:pressurecalc} and find the dimensionless pressure gradient terms in the form of 

\begin{equation}
    -\dfrac{L}{({1}/{2}) \rho U(x)^2} \dfrac{\partial \langle p \rangle}{\partial x} = -\dfrac{L \partial C_p}{\partial x}
\end{equation}

\noindent in terms of the pressure coefficient defined as $C_p ={(\langle p \rangle - p_{\infty})}/({({1}/{2}) \rho U(x)^2}) $ on a horizontal line parallel to the Flat portion of the samples at a height of $y = \pm 0.6 h$ which is at a distance of $0.1 h$ or $\lambda/2$ from the Flat part of the surface on either side, where the 3D effects of the riblets have mostly faded away and it is safe to assume that $\mathbf{u}(x,y,z) = \langle \mathbf{u} \rangle (x,y)$ and $p(x,y,z) = \langle p \rangle(x,y)$. Therefore, we can find the difference between the $\partial \langle p \rangle/\partial x$ and $\partial \langle P^* \rangle/\partial x$, corresponding to terms \raisebox{.5pt}{\textcircled{\raisebox{-.9pt} {2}}} and \raisebox{.5pt}{\textcircled{\raisebox{-.9pt} {3}}} (as well as \raisebox{.5pt}{\textcircled{\raisebox{-.9pt} {4}}} in the LE and LET region), found via the two separate methods, in a dimensionless form as

\begin{equation}
    {\cal G} = \dfrac{-L}{(1/2) \rho U(x)^2}  \left( \dfrac{\partial \langle P*\rangle}{\partial x} - \dfrac{\partial \langle p \rangle}{\partial x} \right) = \dfrac{2mL}{x} + \dfrac{L \partial C_p}{\partial x}
\end{equation}

\noindent and present it for all the riblet samples in figure \ref{fig:p_grad_diff} with the results for the smooth sample shown as reference. (In the LE and LET regions the $\partial/\partial s$ has been transformed to $\partial/\partial x$ using the chain rule $\partial/\partial x =  \partial s/\partial x \ \partial/\partial s$.) As seen in this figure, the non-zero difference between the two terms confirms that the \raisebox{.5pt}{\textcircled{\raisebox{-.9pt} {2}}}, \raisebox{.5pt}{\textcircled{\raisebox{-.9pt} {3}}} and \raisebox{.5pt}{\textcircled{\raisebox{-.9pt} {4}}} have limited non-negligible effects for a finite length sample. Firstly, in the absence of the riblets for the smooth reference, in the LE and LET regions equation \eqref{NSE_average_curve} holds while in the Flat region, equations \eqref{NSE_average2} and \eqref{updated_p1} are simplified to 

\begin{equation}
    \rho \left( u  \dfrac{\partial u }{\partial x} + v  \dfrac{\partial u }{\partial y}  \right) = - \dfrac{\partial P^{*} }{\partial x} + \mu  \dfrac{\partial^2 u }{\partial y^2}  \ \ \ \ \ ; \ \ \ -\dfrac{\partial P^{*} }{\partial x} = - \underbrace{\dfrac{ \partial p }{\partial x}}_{\raisebox{.5pt}{\textcircled{\raisebox{-.9pt} {1}}}} + \underbrace{\mu \dfrac{\partial^2 u }{\partial x^2 }}_{\raisebox{.5pt}{\textcircled{\raisebox{-.9pt} {2}}}}.  %\label{NSE_average2}
\end{equation}

\noindent The viscous term in the streamwise direction , \raisebox{.5pt}{\textcircled{\raisebox{-.9pt} {2}}}, as seen in figure \ref{fig:p_grad_diff} is mostly zero or negative along both sides of the plate in the Flat region. In the leading edge area, the effect of both \raisebox{.5pt}{\textcircled{\raisebox{-.9pt} {2}}} and \raisebox{.5pt}{\textcircled{\raisebox{-.9pt} {4}}} are present and hence $\cal G$ captures its largest deviation from zero and stays mostly negative. Since the smooth surface (of the configuration used here) does not experience any form of flow reversal, we do not expect \raisebox{.5pt}{\textcircled{\raisebox{-.9pt} {2}}} to become positive for this sample and thus the small regions of ${\cal G}>0$ in the LE and LET are most likely due to the contributions from the curvature terms \raisebox{.5pt}{\textcircled{\raisebox{-.9pt} {4}}}. 

Moving to the riblet covered samples, the distribution of $\cal G$ follows a very similar trend as that of the smooth surface, 
and especially for the shallowest riblets of ${\cal R} = 0.5$ family (figure \ref{fig:p_grad_diff}(A)), the distributions are nearly the same with slight differences recorded for the ${\rm Re}_L = 18,500$ cases in the early portion of the Flat regions. Similarly, for all the cases, in the LE region, $\cal G$ is nearly the same as that of the smooth reference, and in the LET region slight differences can be seen for members of ${\cal R} = 1.0$ and $1.5$ families (figure \ref{fig:p_grad_diff}(B) and \ref{fig:p_grad_diff}(C)) which could be due to any of the \raisebox{.5pt}{\textcircled{\raisebox{-.9pt} {2}}}, \raisebox{.5pt}{\textcircled{\raisebox{-.9pt} {3}}}, or \raisebox{.5pt}{\textcircled{\raisebox{-.9pt} {4}}} terms. 

The main differences between the $\cal G$ of the smooth and riblet surfaces in the Flat region is seen among the ${\cal R} = 1.0$ and $1.5$ families. For example, in the early portion of the Flat region of [1.5, 1.5] at all Reynolds numbers (figure \ref{fig:p_grad_diff}(Cc,f,i)) and [1.5, 0.0], [1.5, -1.5], [1.0, 1.0], and [1.0, 0.0] at Reynolds numbers of $18,500$ and $24,200$ (figure \ref{fig:p_grad_diff}(Cd,e,g,h) and \ref{fig:p_grad_diff}(Be,f,h,i)), $\cal G$ is lower than that of the smooth one which is due to the effect of both \raisebox{.5pt}{\textcircled{\raisebox{-.9pt} {3}}} and contributions from the $\mathcal{Z}$ terms. In the latter portion of the Flat region, especially for sample [1.0, 0.0], [1.0, 1.0], [1.5, -1.5], and [1.5, 1.5], unlike the smooth reference, $\cal G$ turns positive for some extent of the plate length. In this region, as the flow inside the grooves has had sufficient distance to develop, it is more likely for the slow down inside the riblets to lead to near stagnant flow inside the grooves with potential for small re-circulation regions \citep{raayai2017drag} turning the viscous terms \raisebox{.5pt}{\textcircled{\raisebox{-.9pt} {2}}} positive, and giving more weight to the contributions of \raisebox{.5pt}{\textcircled{\raisebox{-.9pt}{2}}} than \raisebox{.5pt}{\textcircled{\raisebox{-.9pt}{3}}} here (see more in the upcoming section \ref{sec:n0}). The difference tends to become more visible toward the sharpest textures and also toward samples with larger $\kappa_2$ where the available cross-sectional area within the riblets allows for \raisebox{.5pt}{\textcircled{\raisebox{-.9pt}{2}}} and \raisebox{.5pt}{\textcircled{\raisebox{-.9pt}{3}}} terms to capture larger variations and push ${\cal G}$ to deviate from that of the smooth reference. These cases experience types II, III (all), and IV shear distributions. 

Overall, the similarity of the ${\mathcal{G}}$ of the riblet and smooth samples and the order of magnitude of the difference observed between the two gives us confidence that the spanwise-averaging operation is able to capture a large part of the flow dynamics and is a credible method for extracting valuable information in studying the effect of riblets and the 3D nature of the flow inside their grooves.

\subsubsection{The effective origin, $n_0$} \label{sec:n0}

\begin{sidewaysfigure}
    \centering
    \vspace{37em}
    \includegraphics[width = 1\textheight]{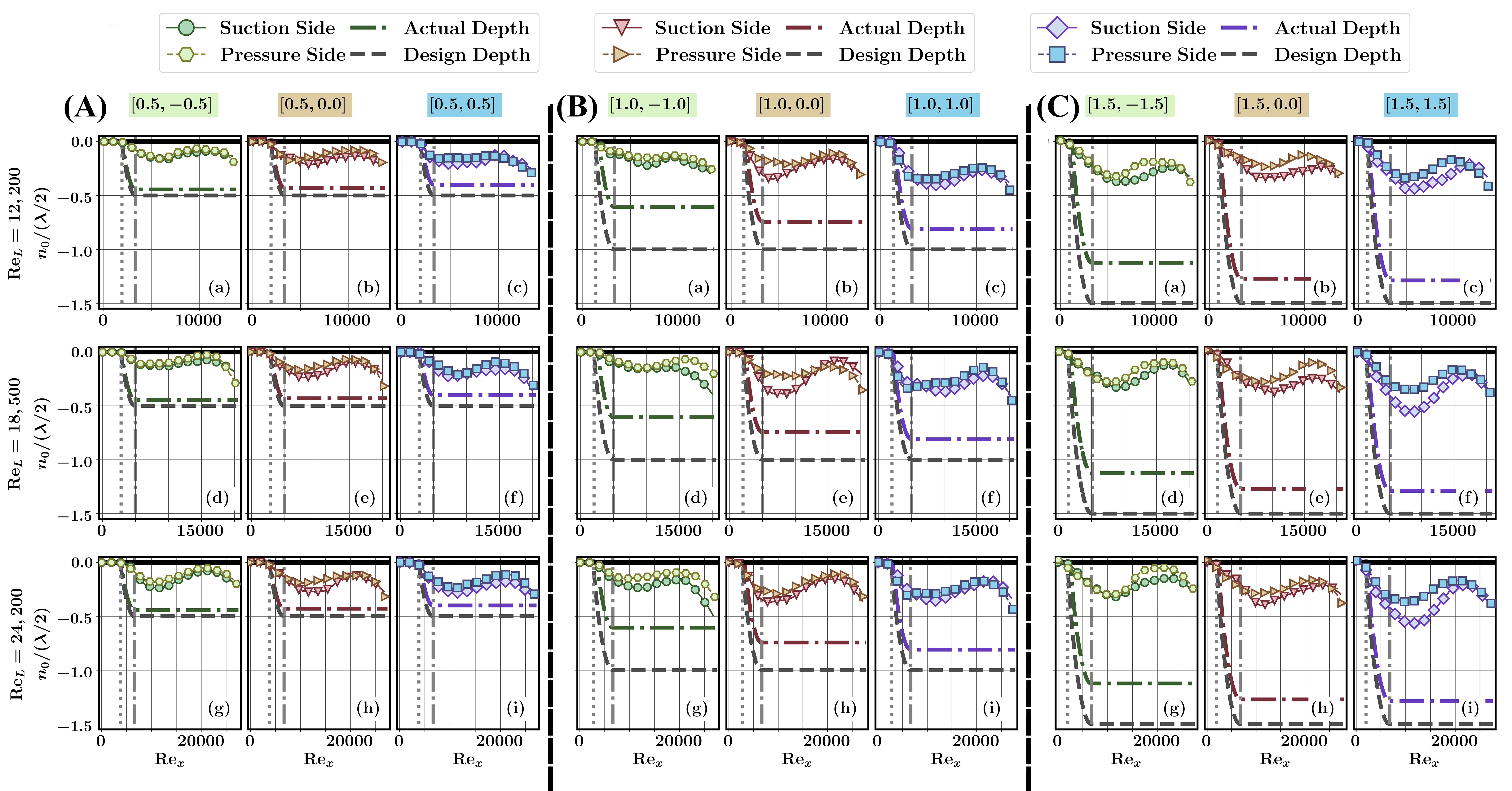}
    \caption{Distribution of the effective origin, $n_0$, for all the riblet samples of (A) ${\cal R} = 0.5$, (B) ${\cal R} = 1.0$, and (C) ${\cal R} = 1.5$ on the suction  and pressure  sides for all the tested Reynolds numbers. Location of the design and measured troughs are also marked on the figures.}
    \label{fig:n0}
\end{sidewaysfigure}

The distribution of the $n_0$ for the suction and pressure sides of all the riblet samples, and for all the tested Reynolds numbers are shown in figure \ref{fig:n0}. In all the plots, the design location of the trough of the riblets and the measured locations are shown for comparison and the $n_0$ values are normalized with $\lambda/2$ which is the more consistent dimensions among all the printed samples. For majority of the cases tested, the distributions of $n_0$, on both sides of the samples, follow a non-monotonic behaviour where initially, following the growth of the riblet height in the LET, the magnitude of the effective origin increases, reaching a maximum in the early portions of the Flat region, before decreasing to a non-zero minimum and afterward again increasing in the vicinity of the trailing edge of the samples until the end of the body. In a few of the cases, namely suction side of [0.5, 0.5] at ${\rm Re}_L = 12,200$ (figure \ref{fig:n0}(Ac)), pressure side of [1.0, -1.0] at ${\rm Re}_L = 12,200$ (figure \ref{fig:n0}(Ba)), and suction side of [1.0, -1.0] at ${\rm Re}_L = 18,500$ (figure \ref{fig:n0}(Bd)), after the initial increase in the magnitude $n_0$, in the Flat region, the effective origin stays nearly constant for a portion of the length of the sample until close to the trailing edge where $n_0$ increases toward the end of the plate. In all the cases, the magnitude of the $n_0$ on the suction sides of the samples are larger then or similar to the pressure sides.

\begin{figure}
    \centering
    \includegraphics[width = 1 \textwidth]{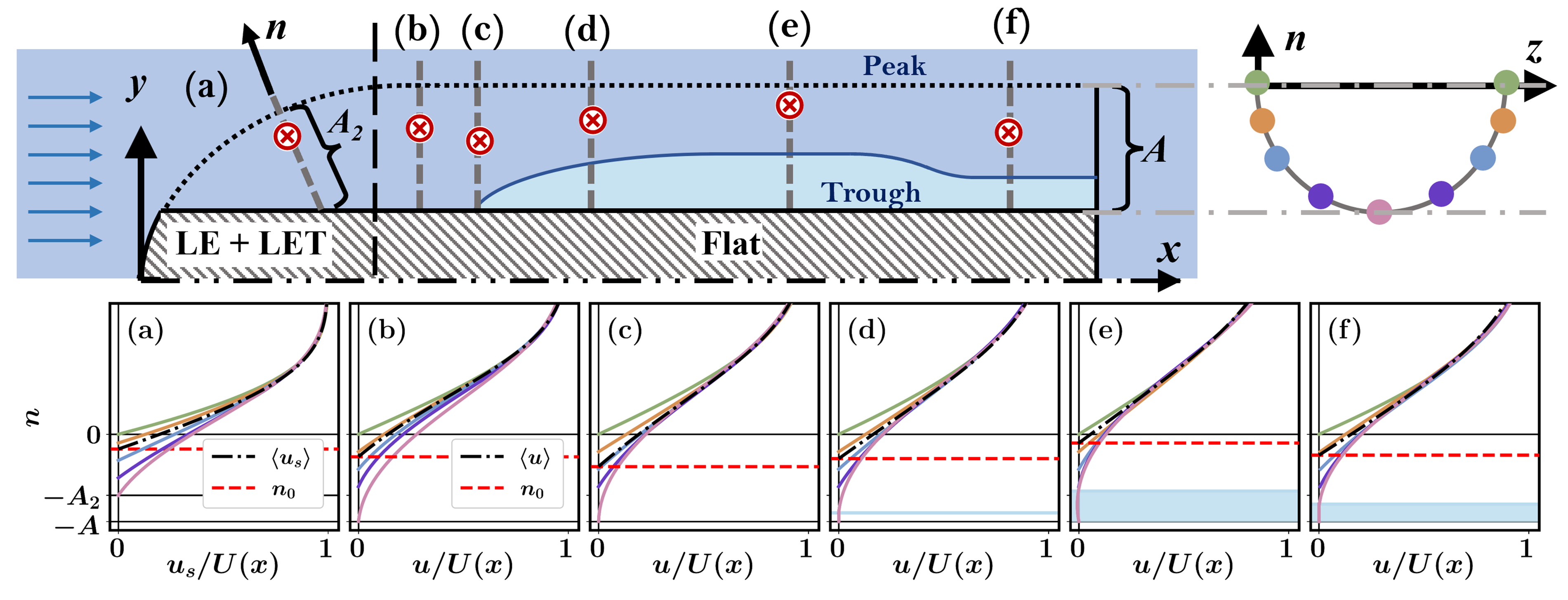}
    \caption{Schematic rendering of the evolution of the effective origin of the velocity profiles (top left) along the grooves of riblet samples for a hypothetical riblet (top right) and the respective local velocity profiles at 5 points (and their mirror images, as shown on the top-right riblet profile with dots with the same colors) in the spanwise direction of the riblet and their respective $\langle u \rangle$ at streamwise locations (a-f) along the sample.}
    \label{fig:BL_evolution}
\end{figure}

Generally, within the ${\cal R} = $ constant families, at a constant Reynolds number, the magnitude of the $n_0$ tends to be larger for $\kappa_2={\cal R}$ (concave) and $\kappa_2 = 0.0$ (triangular) where there is a larger cross-sectional area available in the riblets for the flow to develop compared with the convex ones. Additionally, as the global Reynolds number is increased, we see the absolute values of $n_0$ slightly increasing in the early portion of the Flat region. As it can be expected, as ${\cal R}$ is increased and the available depth of the riblets increases, the effective origin of the velocity profiles is also found to be larger, especially for the concave, $\kappa_2 = {\cal R}$ cases. Among the ${\cal R} = 0.5$ family, for all Reynolds numbers and on both sides, the $n_0$ of the [0.5, 0.0] and [0.5, 0.0] have similar magnitudes in the LET and early portion of the Flat region and toward the second half, the magnitude of the $n_0$ of the [0.5, 0.5] becomes larger than the [0.5, 0.0] sample. Due to the limited resolution of the 3D printer (as listed in table \ref{tbl:geom}), while the height of all the samples are smaller than their design height, the height of the [0.5, 0.5] sample is slightly smaller than the other two samples and thus for this sample the magnitude of $n_0$ can reach to a maximum of between 53-70\%/39-60\% of the measured $A$ in the middle of the suction/pressure side of the riblets. For the [0.5, 0.0] and [0.5, -0.5], the depth of $n_0$ reaches between 50-64\%/38-46\% and 38-53\%/24-41\% of their respective measured $A$ in the middle of the suction (pressure) sides. As the depth of the samples, (i.e. ${\cal R}$) is increased, while we see the absolute value of $n_0$ slightly increasing, the ratio of $\vert n_0 \vert/A$ keeps decreasing. For the family of the ${\cal R} = 1.5$, on the suction/pressure sides the $\vert n_0 \vert /A$ can reach a high as 43\%/33\% as seen for the [1.5, 1.5] sample. Similarly for the ${\cal R} = 1.0$, on the suction/pressure sides, the maximum $\vert n_0 \vert /A$ in the middle of the plate is found to be around 50\%/42\% respectively. As a result, for sharper riblets we expect that a larger volume of the fluid inside the grooves to be in near stagnant condition compared to the shallower grooves where the effective origin can penetrate more than 50\% of the height of the riblet.

The trend in the effective origin, $n_0$ of the spanwise-averaged velocity profiles in the streamwise direction can be explained further with the idea of flow retardation inside the grooves that leads to the creation of a layer of stagnant (or slow-moving) fluid (figure \ref{fig:BL_evolution}) which has been shown in previous numerical simulations \citep{raayai2017drag, chu1993direct, choi1993direct}. Here, as the BL develops along the plate and inside the riblets, the magnitude of the effective origin of the $\langle u \rangle$ (or $\langle u_s \rangle$) profiles slowly increases until reaching a maximum. From this point, first a layer of slow-moving fluid starts to develop inside the grooves and grows along the length of the plate. This layer due to its near-zero velocity does not communicate with the rest of the flow and acts as a blockage, pushing the moving fluid to the outside, effectively \emph{squeezing} the flow between the higher $n_0$ and the inviscid outer flow and at times leading to a faster rate of increase in $m$. This moves the location of $n_0$ outward resulting in a decrease in the magnitude of the $n_0$. This continues until close to the vicinity of the trailing edge where the velocity in the $y$ direction, $v$, outside the grooves changes direction toward the sample (instead of away from the sample, see figure \ref{fig:shear_type}(f)) and effectively pushes the flow inside the grooves and increases the magnitude of $n_0$. In this region, the flow is more attached to the surface with larger $m$ values (as seen in section section \ref{sec:m}), and also larger values of $\langle C_f \rangle (x)$ than the earlier parts of the Flat region where the magnitude of $n_0$ reaches a local maximum (as discussed in section \ref{sec:shear}). 

In some cases, the quiescent flow layer can lead to a slight re-circulation in the flow \citep{raayai2017drag}, which can result in the viscous diffusion terms in \raisebox{.5pt}{\textcircled{\raisebox{-.9pt} {2}}} of equation \eqref{updated_p1} to  become positive, leading to ${\cal G}>0$ in the latter part of the sample as seen in section \ref{sec:m}. Comparing figures \ref{fig:n0} and \ref{fig:p_grad_diff}, one can see that for cases with ${\cal G}>0$ areas in the Flat region, the ${\cal G}>0$ region is very close to where the magnitude of the $n_0$ reaches a local minimum. In this region, with a very slow moving fluid inside the grooves, the contribution from term \raisebox{.5pt}{\textcircled{\raisebox{-.9pt} {3}}} is likely very small and ${\cal G}$ becoming even slightly positive can serve as an indication of the potential for existence of flow reversal in these cases.  

The location of the origin of the normal coordinate is a complicating matter for the experimental efforts in the analysis of the flow over riblet surfaces. \citet{wallace1988viscous} chose the geometric average of the height of the peak and trough of the riblets (midway) as the origin for their analysis. Later as discussed in the section \ref{intro}, the protrusion height model was introduced \citep{luchini1991resistance, luchini1995asymptotic, bechert1997experiments, gruneberger2011drag} as the origin of the velocity profiles below the level of the grooves. Mainly calculated using a Stokes flow analysis inside the grooves, the difference between the location of the protrusion heights seen by the streamwise and spanwise motions have been used to find correlation for the drag reduction values in turbulent flows \citep{bechert1997experiments, wong2024viscous, garcia2011drag}. Here, with the finite length of the sample and the variations expected in the streamwise direction, as well as the laminar nature of the flow, we expect the distribution of $n_0$ to also vary along the length and thus instead of using a Stokes flow approach or a linear estimation, we use the fitting process to extract the effective origin.

There are similarities between the idea of the effective origin and the other definitions, and overall, locally the velocity profiles of shear-reducing riblets experience a lower $n_0$ (higher magnitude) than shear-increasing ones. For example, in the case of [1.5, 1.5] at ${\rm Re}_L = 24,200$ and on the suction sides, the available cross-sectional area of the grooves, makes it possible for the origin of the velocity profile at ${\rm Re}_x = 10,000$ to reach 43\% of the texture height (56\% of the half-spacing) and thus allow the $\langle u \rangle$ to take a more detached form, and $m$ to get as low as -0.057, $\cal G$ to get to visibly lower values compared to the smooth (see figure {\ref{fig:p_grad_diff}}(Ci)), and ultimately close to 30\% local shear reduction (figure {\ref{fig:shear}(Cc)}). In turn on the pressure side of [1.0, 0.0] at ${\rm Re}_L = 12,200$, $n_0$ is able to reach to as low as 28\% of the measured height (21\% of the half-spacing) at ${\rm Re}_x = 6,000$, where the velocity profile is more attached compared to the smooth reference with $m = 0.003$, and we see a similar ${\mathcal{G}}$ compared with the smooth reference (figure {\ref{fig:p_grad_diff}(Cc))}, and recording about 10\% shear increase (figure {\ref{fig:shear}(Cc))}. Ultimately, as demonstrated, for a limited size body, the effect of the $n_0$ and $m$ on the frictional shear/drag changes are more intertwined where the available cross-sectional area inside the riblets, the Reynolds number, and the pressure distribution all guide the development of the BL and additional work needs to be done to translate these two parameters into predictive tools.

\subsection{Pressure Drag} \label{sec:pressureDrag}

Somewhere between 23-36\% of the total drag experienced by the samples is attributed to the $D_{p}$. The pressure drag is due to the finite size of the sample and the resulting pressure distribution around the entire body. Independent of the total skin friction drag, the pressure drag also experiences alterations as a result of the presence of the riblets. This drag component is found cumulatively with the friction drag using a control volume analysis where the sum of the friction and pressure drag can be found as a total reaction force applied to the sample, $D_{\rm CV}$, also known as the profile drag \citep{fu2023doublelightsheet}. (To avoid confusion between the profile and pressure drag, here we use a generic $D_{\rm CV}$, to represent the force calculated via the control volume method.) Therefore, $D_{p} = D_{\rm CV} - D_{f}$. In summary, we use the Reynolds Averaged Integral Momentum (RAIM) formulation \citep{Ferreira_2021, fu2023doublelightsheet, suchandra2023impact} 

\begin{equation}\label{eq:RAIM}
\mathbf{D}_{\rm CV} \, = \, - \, \rho \int \overline{(\mathbf{u} + \mathbf{u'}) [(\mathbf{u} + \mathbf{u'}) \cdot \mathbf{n}_S]} dS \,  - \, \int p \, \mathbf{n}_S \, dS
\end{equation}

\noindent where $\mathbf{n}_S$ is the normal to the control surface and $S$ is the area of the control surface. We also assume that we are far enough from the riblets that the 3D nature of the velocity profile has subsided. We place the boundaries of the control volumes on a (1) plane prior to the leading edge at the earliest possible available location at around $x/L \approx -0.35$, (2,3) two parallel planes far from the BL on either side of the sample at $y/L = \pm 0.12$, and (4) a plane after the trailing edge, at $x_{_{\rm TE}}> L$ (similar to the procedure followed by \citet{fu2023doublelightsheet}). We fix the first 3 boundaries and use 40 control volumes with 40 different $x_{_{\rm TE}}$ to calculate the $D_{\rm CV}$ for all the control volumes and present the mean of the values and their 95\% confidence intervals in figure \ref{fig:CfCpC3D}. Here, we assume at the earliest available point before the leading edge ($x/L \approx -0.35$) and far from the BL ($y/L = \pm 0.12$) the pressure is $p_{\infty} = 0$ and perform the directional integration following the procedure discussed in section \ref{sec:pressurecalc} to find the pressure distribution needed in equation \eqref{eq:RAIM}. 

\begin{figure}
    \centering\includegraphics[width = 1 \textwidth]{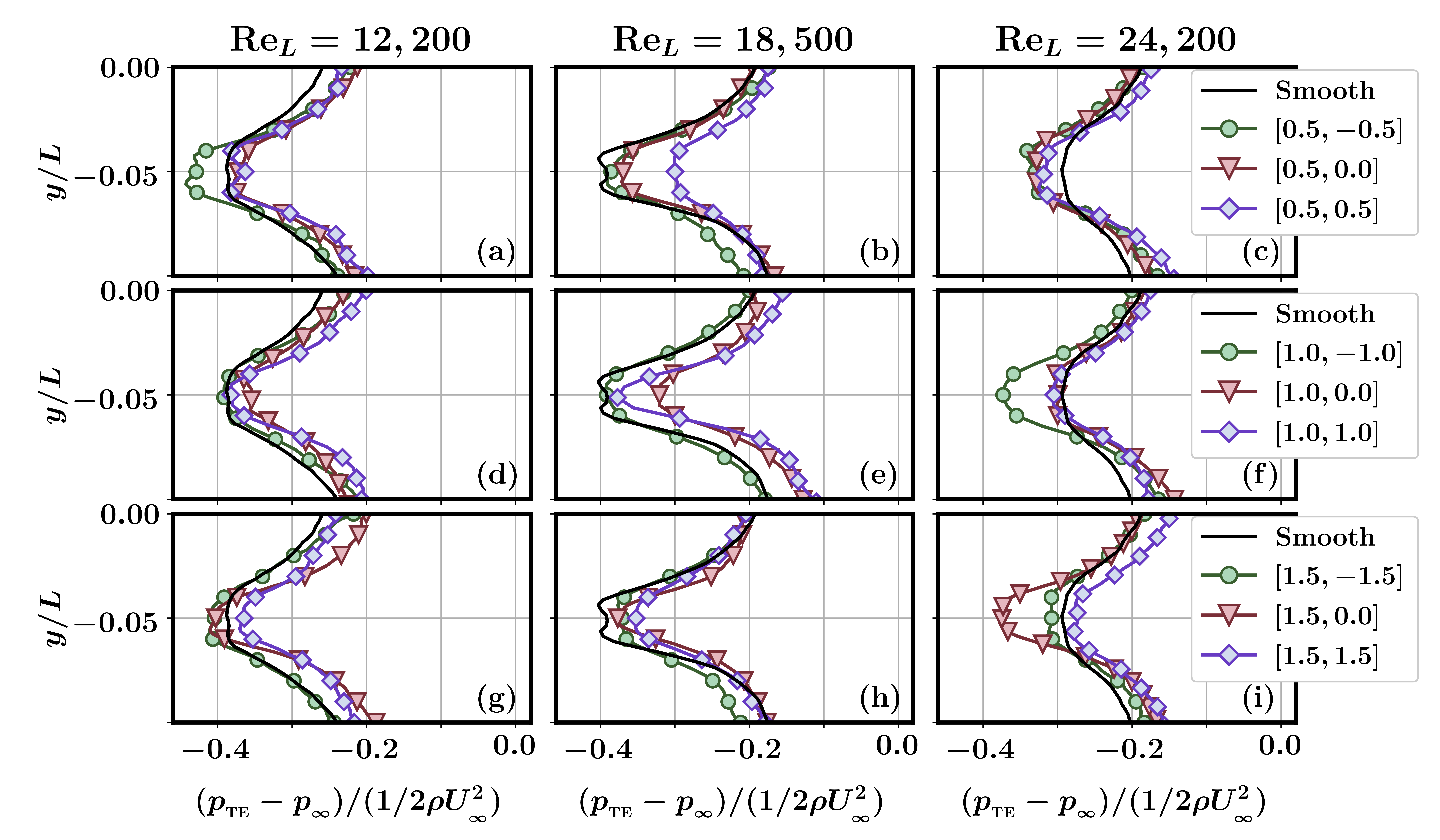}
    \caption{Distribution of pressure past the trailing edge of the plates at $x/L = 1.05$ for families of (a,b,c) ${\cal R} = 0.5$, (d,e,f) ${\cal R} = 1.0$, and (g,h,i) ${\cal R} = 1.5$ for three global Reynolds numbers. The pressure distribution of the smooth reference is shown with solid black line.}
    \label{fig:TE_pressure}
\end{figure}

In the presence of the riblets, samples only experience a marginal difference in the momentum distribution crossing the control volume boundaries (1) and (4) at the leading and trailing edges, while the pressure difference between the two planes, $p_{_{\rm TE}} - p_{_{\rm \infty}}$ (as shown in figure \ref{fig:TE_pressure}), experiences a clear difference between the riblet-covered samples and the smooth reference. For the samples experiencing reductions in the pressure drag, this difference comes out in the form of pressure recoveries past the trailing edge (see figure \ref{fig:TE_pressure}) which ultimately enhances the overall drag reduction of the samples. 

\begin{sidewaysfigure}
    \centering
    \vspace{37em}
    \includegraphics[width = 1\textheight]{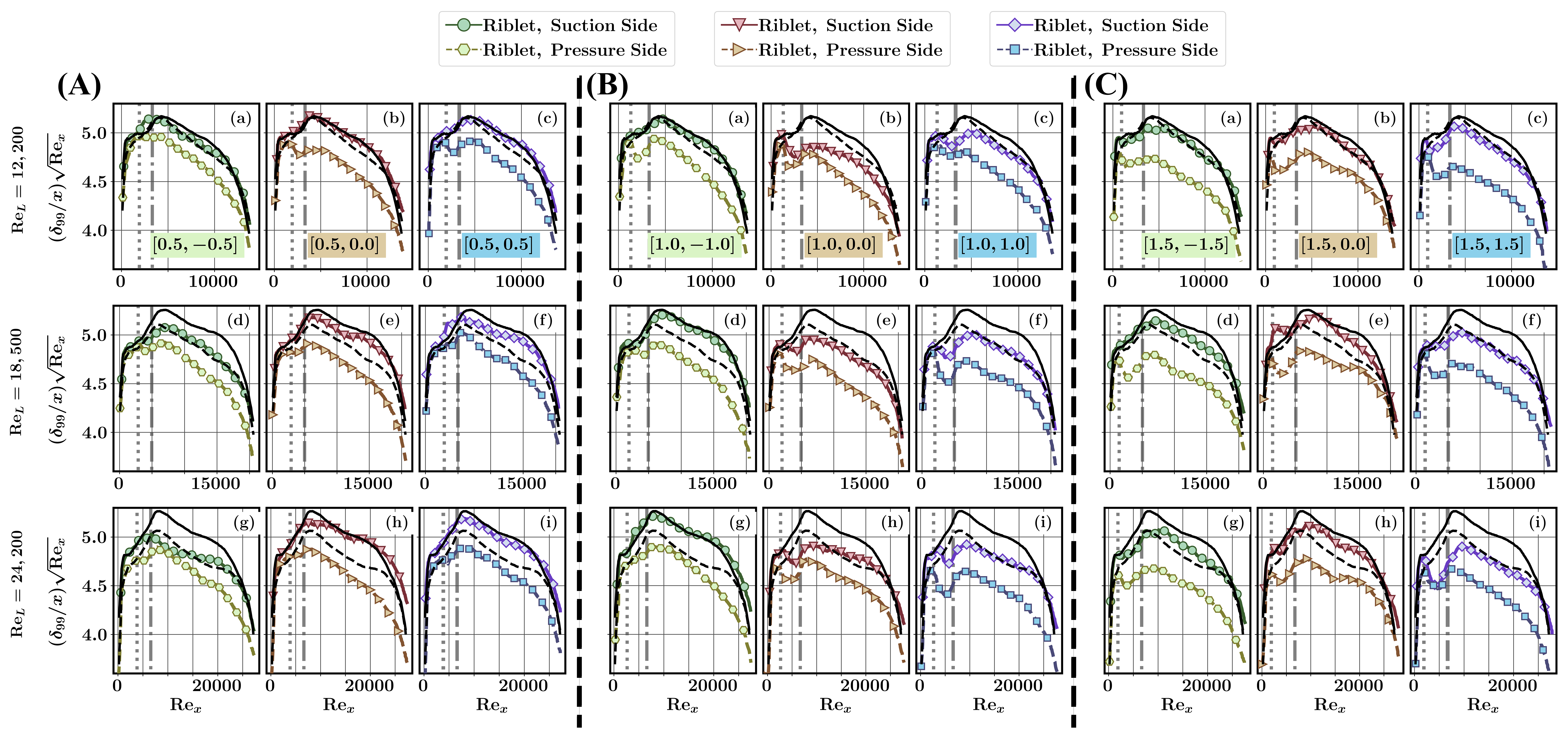}
    \caption{Distribution of the BL thickness, $\delta_{99}$, normalized by local $x/\sqrt{{\rm Re}_x}$, for all the riblet samples of (A) ${\cal R} = 0.5$, (B) ${\cal R} = 1.0$, and (C) ${\cal R} = 1.5$ on the suction  and pressure  sides for all the tested Reynolds numbers. The $\delta_{99}$ of the smooth reference on the suction and pressure sides are shown with solid and dashed black lines respectively.}
    \label{fig:delta}
\end{sidewaysfigure}

Riblets affect the near wall BL which results in differences in the $\langle u \rangle$ of the riblet samples compared to the smooth reference. On the one hand, we have non-zero $\Delta \langle \tau_{\rm w} \rangle$ which cumulatively affects the skin friction portion of the total drag. On the other hand, the changes in the velocity profiles affect the location of the edge of the BL and hence the BL thickness. Here, we calculate the BL thickness in terms of $\delta_{99}$ where $u(x,\delta_{99},z) = \langle u \rangle (x,\delta_{99})= 0.99 U(x)$ (and equivalently in the leading edge $u_s(x,\delta_{99},z) = \langle u_s \rangle (x,\delta_{99},z) = 0.99 U(x)$) and plot the results for suction and pressure sides of all the cases in figure \ref{fig:delta}. The $\delta_{99}$ values are with respect to the $n=0$ or the location of the peak of the riblets, which is the same as the location of the boundary of the smooth sample and are normalized by $x/\sqrt{{\rm Re}_x}$. As it can be seen in figure \ref{fig:delta}, in presence of the riblets and with the available space inside the grooves for the velocity profiles to develop, the thickness of the BL on both sides are smaller than the case of the smooth reference. Especially with larger $m$ on the pressure sides, the BL thickness on the pressure side is even smaller than the suction side. Thus, overall, the edge of the BL as seen by the inviscid flow is different and also slightly thinner for the riblet samples compared with the smooth reference which can lead to lower pressure drops at the trailing edge of the samples as well (similar behaviour has been reported in the numerical simulations of \citet{mele2020effect}).

Lastly, this reduction in the overall thickness of the fictitious boundary of the sample and the BL seen by the outer inviscid flow can be a substantial help to samples that cannot capture a reduction in the frictional drag compared to the smooth surface. Example of that is the [1.0, 0.0] sample which from a frictional point is not able to reduce the drag force, however, as seen in figure \mbox{\ref{fig:m}(Bb,e,h)}, with a large extent of the plate experiencing $m$ values larger than the smooth reference, the $\delta_{99}$ of this sample experiences enough reduction on both sides, especially in the early part of the Flat region (see figure \ref{fig:delta}(Bb,e,h)) to experience noticeable levels of pressure recovery (figure \ref{fig:TE_pressure}(b,d,e)) at the trailing edge and ultimately becoming drag-reducing (figure \ref{fig:CfCpC3D}).

\section{Conclusions} \label{sec:conclusion}

Here, we evaluate the possibility of using riblets as a drag-reducing technique on a standalone, finite-sized slender body. We use a 3-tiered measurement approach to capture the total drag force and decompose it into the frictional and pressure components. We further use the local shear stress and pressure and the distribution of the velocity profiles to explore the drag-changing performance of the riblets.

Overall, a majority of the cases presented here, operated in high Reynolds number laminar regimes, showed some level of drag reduction (total drag), with drag-increasing cases all having a convex riblet shape (${\kappa_2} = -{\cal R}$). The largest total reduction is seen for the [1.0, 1.0] sample at ${\rm Re}_L = 18,500$ and $24,200$ with 5.8-6.5\% reduction recorded. Due to the finite-size of the sample, between 42-52\% of the drag is due to the frictional component, and around 23-34\% due to the pressure, with the remaining 20-24\% attributed to other effects which cannot be captured via planar PIV measurements. The contribution of the frictional drag to the total drag change is limited, and as the global Reynolds number is increased, more of the cases experienced a reduction in the $D_f$. On the other hand, the contribution of the pressure drag to the reduction is more pronounced with most of the cases experiencing some level of reduction in the $D_p$ compared to the smooth reference. 

Further investigation of the localized shear stress distribution reveals a more complex behaviour that leads to the subpar performance of the riblets in terms of the frictional drag force. On the one hand, the asymmetry of the flow field, due to the angle of attack of the sample, results in an asymmetry in the contribution of the suction and pressure side of the sample to the $D_f$. Suction sides of most samples are drag-reducing while the pressure sides tend to be drag-increasing and thus cumulatively the increase of the pressure side counteracts the gains of the suction side. Only 4 of the cases (see figure \ref{fig:phase_map}) see frictional reductions on both sides. On other other hand, the presence of riblets gives rise to 4 different $\langle C_f \rangle (x)$ patterns (Type I-IV). While the pattern of Type I is similar to that of the smooth reference, with different rates compared to the smooth one where $\langle C_f \rangle (x)$ takes a decreasing trend along the length, the patterns of Types II-IV show clear differences leading to $\langle C_f \rangle (x)$ becoming either constant or increasing along the length of the sample. This results in a non-monotonic distribution of the $\Delta \langle \tau_{\rm w} \rangle$ with the possibility of having regions of both $\langle C_f \rangle(x) > C_{f,0}(x)$ and $\langle C_f \rangle(x) < C_{f,0}(x)$ (i.e. $\Delta \langle \tau_{\rm w} \rangle >0$ or $\Delta \langle \tau_{\rm w} \rangle <0$) where again the reductions are counteracted by the increases, leading to the frictional drag integral capturing much less reduction in the frictional drag than the riblets' local shear-reducing potential observed. 

With access to the $\langle C_f \rangle (x)$ distribution on both sides of the sample, we also decompose the drag into components within the leading edge (LE and LET) and Flat segments of the samples. We see that the choice of mimicking the shark nose in the design of the leading edge allows us to avoid large $\Delta \langle \tau_{\rm w} \rangle$ values in that region, and the incremental growth of the riblets also guides the development of the velocity field and $\langle C_f \rangle (x)$ in the LET region. In this design the $\langle C_f \rangle (x)$ of the LE region is not available for modification. 

The impact of riblets on the flow field is mostly confined to the BL near the wall and we use this idea in our experimental procedure to measure the spanwise-averaged velocity field around the body using 2D-2C PIV. By fitting the velocity profiles \emph{locally} to an updated form of the FS family of BL solutions, we find the parameters $m$ (capturing the effect of pressure gradient, streamwise and spanwise viscous diffusion, nonlinear terms due to span-wise averaging, and curvature of leading edge), and $n_0$ (effective origin of the velocity profiles). We see similar, repeated patterns in the distributions of $m$ and $n_0$, while the magnitudes of the two parameters show an intertwined relationship which is driven by the flow dynamics and the available cross-sectional space inside the grooves. The distribution of $n_0$ shows clear signs of flow retardation and creation of layers of slow moving fluid inside the grooves which then leads to the flow being pushed out of the grooves and magnitude of the $n_0$ decreasing prior to the trialing edge. This directly impacts the distribution of $m$ where after an initial decrease, $m$ takes an increasing trend in the second half of the plate, with rates faster than that of the smooth reference (as the flow is squeezed between the higher $n_0$ and the edge of the BL) and thus resulting in  Types II-IV shear distributions. 

In addition, we use the distribution of $m$ alongside the distribution of the pressure gradient on a line parallel to the wall, in the form of the ${\cal G}$, to explore the effect of the streamwise and spanwise viscous terms, nonlinear terms due to spanwise averaging, and the curvature effects. While $m$ cannot distinguish between the order of magnitude of these terms, based on the physics, we estimate that the BL development along the grooves and inside the slow moving fluid layer leads to re-circulation regions as identified by ${\cal G}>0$ regions. Overall, the similarity of the ${\cal G}$ of the smooth and riblet surfaces gives us confidence that the effect of the spanwise-averaging on the nonlinear terms are minimal and we can extract valuable information from planar PIV experiments for riblet surfaces. 

The impact of riblets on the near wall boundary layer is also evident in the changes in the location of the edge of the BL as seen by the inviscid outer flow. The cross-sectional space available inside the grooves and the distribution of $m$ along the side of the samples leads to, on average, thinner BLs compared to the smooth reference, especially on the pressure sides of the samples. This results in a pressure recovery seen in the trailing edge of most of the riblet samples and thus clear reductions in the pressure drag compared to the smooth reference. This not only helps the samples which were $D_f$-reducing, but makes it possible for a few of the $D_f$-increasing samples to become cumulatively drag-reducing when the pressure component is added. 

Overall, the evidence presented supports the idea that the implementation of riblet for reducing the drag on finite-sized, stand-alone samples is feasible. However, to gain the most benefit from this approach, it requires a more comprehensive design plan which considers the effect of the friction and pressure drags simultaneously. While in the ideal design, the riblets on both sides follow a Type I $\langle C_f \rangle (x)$ distribution along the length, are cumulatively $D_f$-reducing, and the edges of the BL are sufficiently adjusted for the $D_p$ to also be reduced, there are a variety of ways where the riblet design can be optimized to enhance the reduction in either $D_f$ or $D_p$ to achieve desired levels of reductions. The leading and trailing edges of the body play important roles in guiding the development of the $\langle C_f \rangle (x)$ along the length of the sample and potential adjustments to the curvature of those regions could improve the performance of the riblets as shear-reducing agents. In addition, the reduction in the pressure component of the drag due to the adjustment in the BL thickness opens up another avenue for riblets to also be considered for bulkier vehicles which experience a larger contribution from the pressure drag. 

Further modelling efforts supported by numerical simulations or stereo- or tomo-PIV efforts focused on the characterization of the flow inside the grooves can be used to add additional validation to the spanwise-averaging method presented here and improve our predictive capability. In addition, extensions of this work to larger Reynolds numbers, turbulent flows, as well as those with non-constant pressure distributions will be instrumental in expanding the use of drag-reducing riblets to smaller and bulkier vehicles as well.  

\backsection[Acknowledgements]{The optical profilometry has been performed at the Harvard University Center for Nanoscale Systems (CNS); a member of the National Nanotechnology Coordinated Infrastructure Network (NNCI), which is supported by the National Science Foundation under NSF award no. ECCS-2025158.}

\backsection[Funding]{This research has been supported by the Rowland Fellows program at Harvard University.}

\backsection[Declaration of interests]{The authors report no conflict of interest.}

\backsection[Data availability statement]{The raw data that support the findings of this study are available from the corresponding author upon reasonable request.}

\backsection[Author ORCIDs]{S. Fu, https://orcid.org/0000-0003-4768-3702; S. Raayai-Ardakani, https://orcid.org/0000-0002-1733-3309}

{\backsection[Author contributions]{S.F.: Investigation, Formal analysis, Visualization, Writing - original draft, Writing - review \& editing. S.R.: Conceptualization, Methodology, Formal Analysis, Software, Visualization, Writing - original draft, Writing - review \& editing, Supervision, Funding acquisition.}}

\appendix

\section{Local control volume analysis inside the grooves for obtaining spanwise-averaged wall shear stress}\label{appA}

\begin{figure}[b]
    \centering
    \includegraphics[width=1\linewidth]{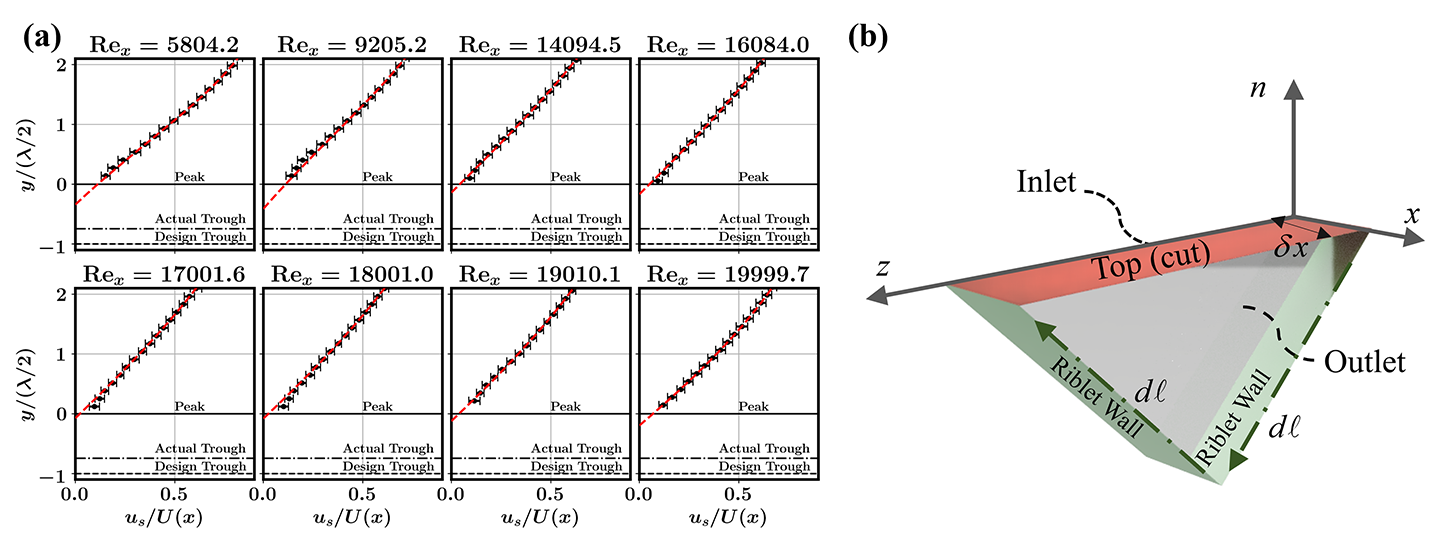}
    \caption{(a) Distribution of the tangential velocity profiles and their 95\% confidence intervals, at 8 different location along the suction side of the [1.0, 0.0] sample at ${\rm Re}_L = 18,500$, and the FS fits to the profiles and the extrapolations below the peak levels showing the location of $n_0$. Dashed and dash-dotted black lines correspond to the design and measured location of the troughs respectively. (b) A control volume inside the riblets, cut at the peak of the grooves with a thickness of $\delta  x$ in the $x$ direction. }
    \label{fig:Vel_CV}
\end{figure}

Since we do not have access to the velocity profiles inside the grooves to capture the shear stress distribution, we use a simple control volume, bounded by the riblet wall on the bottom, and cut at the peak level of $n=0$, with an infinitesimal depth of $\delta x$, such as the one shown in Fig. \ref{fig:Vel_CV}(b). Using this, we can write 

\begin{equation}
    \begin{split}
    &-\int_x^{x+\delta x} \int_0^{\lambda} \tau_{_{n=0}}(x,n=0, z) dz dx  + \int_x^{x+\delta x} \int_{\rm riblet} \tau_{\rm w} d \ell dx + \\ & \int_{\rm inlet/Outlet} p(x) - p(x+\delta x) dS =  \sum_{i} \int_{S_i} \rho \mathbf{u} (\mathbf{u} \cdot {\mathbf{n}}_S) dS_i 
     \end{split}
\end{equation}

\noindent where $\tau_{n=0}$ is the shear stress distribution on the top boundary which is included instead of the cut, $\tau_{\rm w}$ is the shear stress distribution on the riblet wall, $p$ is the pressure, and $\mathbf{n}_S$ is the unit normal to the wall of the boundaries of control surfaces, and $i \in {\rm [Top, Inlet, Outlet, Riblet]}$. For this control volume, at the limit of $\delta x \rightarrow 0$ on the top boundary at $n=0$

\begin{equation}
    \int_{x}^{x+\delta x} \int_0^{\lambda} (\rho uv) dz dx \approx 0
\end{equation}

\noindent and we also assume that with the slow down of the flow inside the grooves, between the inlet and outlet the variations in the velocity and pressure inside the grooves are also very small $u(x;y,z) \approx u(x+\delta x;y,z)$ and $p(x;y,z) \approx p(x+\delta x;y,z)$ and 

\begin{equation}
\begin{split}
    &\int_{{\rm inlet}, \ x} (\rho u^2) dS \approx \int_{{\rm outlet}, \ x+\delta x} (\rho u^2) dS \\ 
    & \int_{\rm inlet/Outlet} p(x) - p(x+\delta x) dS \approx 0
\end{split}
\end{equation}

\noindent Thus, 

\begin{equation}
    -\int_x^{x+\delta x} \int_0^{\lambda} \tau_{n=0}(x,z) dz dx  + \int_x^{x+\delta x} \int_{\rm riblet} \tau_{\rm w} d \ell dx = 0
\end{equation}

\noindent and 

\begin{equation}
    \langle \tau_{\rm w} \rangle(x) = \dfrac{\int_0^{\lambda} \tau_{n=0}(x,z) dz}{\lambda} \approx \dfrac{\int_{\rm riblet} \tau_{\rm w}(\ell) d \ell}{\lambda}.
\end{equation}

\noindent Hence, just by having access to the average shear stress on the plane at the $n=0$, we can evaluate the spanwise-averaged shear stress experienced by the riblet surface. 

\section{Spanwise-averaged Navier-Stokes equations in the Flat region}\label{appB}

In the Flat region of the riblet samples, we use the Cartesian form of the Navier-Stokes equation in the $x$ direction and apply the spanwise-averaging operation defined as 

\begin{equation}
    \langle \ldots \rangle = \dfrac{1}{\lambda} \int_0^{\lambda} \ldots \ dz
\end{equation}

\noindent to get

\begin{equation}
    \rho \left( \left \langle u \dfrac{\partial u}{\partial x} \right \rangle + \left \langle v \dfrac{\partial u}{\partial y} \right \rangle + \left \langle w \dfrac{\partial u}{\partial z} \right \rangle \right) = - \left\langle \dfrac{\partial p }{\partial x} \right\rangle + \mu \left( \left\langle\dfrac{\partial^2 u}{\partial x^2} \right\rangle + \left\langle\dfrac{\partial^2 u}{\partial y^2} \right\rangle + \left\langle \dfrac{\partial^2 u}{\partial z^2} \right\rangle  \right)
\end{equation}

\noindent The derivatives in the $x$ and $y$ direction and the averaging operation can commute thus

\begin{equation}
    \rho \left( \left \langle u \dfrac{\partial u}{\partial x} \right \rangle + \left \langle v \dfrac{\partial u}{\partial y} \right \rangle + \left \langle w \dfrac{\partial u}{\partial z} \right \rangle \right) = - \dfrac{\partial  \langle p \rangle }{\partial x} + \mu \left( \dfrac{\partial^2 \langle u \rangle}{\partial x^2} + \dfrac{\partial^2 \langle u \rangle}{\partial y^2} + \left\langle \dfrac{\partial^2 u}{\partial z^2} \right\rangle  \right). \label{NSE_part1}
\end{equation}

\noindent Using the product rule
% \begin{equation}
\begin{align}
    \left \langle u \dfrac{\partial u}{\partial x} \right \rangle = & \dfrac{\partial \langle u u \rangle}{ \partial x} - \left \langle u \dfrac{\partial u}{\partial x} \right \rangle \label{PR1}\\
    \left \langle v \dfrac{\partial u}{\partial y} \right \rangle = & \dfrac{\partial \langle u v \rangle}{ \partial n} - \left \langle u \dfrac{\partial v}{\partial y} \right \rangle \label{PR2}\\
    \left \langle w \dfrac{\partial u}{\partial z} \right \rangle = & \left \langle  \dfrac{\partial (u w) }{ \partial z} \right \rangle- \left\langle u \dfrac{\partial w}{\partial z} \right\rangle. \label{PR3}
\end{align}

\noindent Due to the periodicity of the velocity in the $z$ direction, $u(z=\lambda) = u(z=0)$ and $w(z=\lambda) = w(z=0)$ and we get

\begin{equation}
    \left \langle \dfrac{\partial uw}{\partial z} \right \rangle = \dfrac{1}{\lambda} \int_0^\lambda  \dfrac{\partial uw}{\partial z}  dz = \dfrac{1}{\lambda} \left( uw \bigg\vert_{z=0} - uw \bigg \vert_{z=\lambda} \right)  = 0 \label{uw_0}
\end{equation}

\noindent and using equations \eqref{PR1}, \eqref{PR2}, \eqref{PR3}, and continuity, the left hand side of equation \eqref{NSE_part1} is simplified to

\begin{equation}
    \rho \left( \left \langle u \dfrac{\partial u}{\partial x} \right \rangle + \left \langle v \dfrac{\partial u}{\partial y} \right \rangle + \left \langle w \dfrac{\partial u}{\partial z} \right \rangle \right) = \rho \left(\dfrac{\partial \langle uu \rangle}{\partial x} + \dfrac{\partial \langle uv  \rangle}{\partial y} \right).
\end{equation} 

\noindent On the right hand side, with symmetry and periodicity, for $\vert y \vert >h/2$ (see figure \ref{fig:averaging_BC}(a)), and the fact that outside the riblets, above the peaks, the velocity gradient needs to be continuous everywhere, we have

\begin{equation}
    \dfrac{\partial u}{\partial z} \bigg \vert_{z=\lambda^+}  = \dfrac{\partial u}{\partial z} \bigg \vert_{z=\lambda^-} = \dfrac{\partial u}{\partial z} \bigg \vert_{z=0^+} = \dfrac{\partial u}{\partial z} \bigg \vert_{z=0^-} = 0
\end{equation}

\noindent resulting in 

\begin{equation}
    \left\langle \dfrac{\partial^2 u}{\partial z^2} \right\rangle = \dfrac{1}{\lambda} \int_0^{\lambda} \dfrac{\partial^2 u}{\partial z^2} dz = \dfrac{1}{\lambda} \left(\dfrac{\partial u}{\partial z} \bigg \vert_{z=\lambda} - \dfrac{\partial u}{\partial z} \bigg \vert_{z=0} \right) = \dfrac{-2}{\lambda} \dfrac{\partial u}{\partial z} \bigg \vert_{z=0} = 0 \label{u_z_0}
\end{equation}

\noindent Inside the grooves, for $y_{\rm trough} \leq \vert y \vert  \leq h/2$, where the walls start at a later $z_{\rm w1}\geq 0$ and terminate at an earlier $z_{\rm w2} \leq \lambda$ (where $\lambda - z_{\rm w2} = z_{\rm w1}$, see figure \ref{fig:averaging_BC}(a)), with symmetry we have

\begin{equation}
    \left\langle \dfrac{\partial^2 u}{\partial z^2} \right\rangle = \dfrac{1}{\lambda} \int_0^{\lambda} \dfrac{\partial^2 u}{\partial z^2} dz = \dfrac{1}{\lambda} \left(\dfrac{\partial u}{\partial z} \bigg \vert_{z=z_{\rm w2}^-} - \dfrac{\partial u}{\partial z} \bigg \vert_{z=z_{\rm w1}^+} \right) = \dfrac{-2}{\lambda} \dfrac{\partial u}{\partial z} \bigg \vert_{z=z_{\rm w1}^+} \label{u_z_02}
\end{equation}

\noindent Therefore, equation \eqref{NSE_part1} is written as

\begin{equation}
    \rho \left(\dfrac{\partial \langle uu \rangle}{\partial x} + \dfrac{\partial \langle uv  \rangle}{\partial y} \right) = - \dfrac{\partial \langle p \rangle }{\partial x} + \mu \left( \dfrac{\partial^2 \langle u \rangle}{\partial x^2} + \dfrac{\partial^2 \langle u \rangle }{\partial y^2} \right) + {\mathcal{Z}}
\end{equation}

\noindent with 

\begin{equation}
    {\mathcal{Z}} = \left\{
    \begin{alignedat}{3} 
        & - \dfrac{2\mu}{\lambda} \dfrac{\partial u}{\partial z} \bigg \vert_{z=z_{\rm w1}^+}  \ \ \ \ \ & y_{\rm trough} \leq \vert y \vert  \leq h/2 \\
        &0 \ \ \ \ \ &\vert y \vert  > h/2.
    \end{alignedat}
    \right.  \label{viscousBCZ}
\end{equation}

\noindent Ultimately we re-write the equations in the form

\begin{equation}
    \rho \left(\langle u \rangle \dfrac{ \partial \langle u \rangle}{\partial x} + \langle v  \rangle \dfrac{\partial \langle u  \rangle}{\partial y} \right) = - \dfrac{\partial \langle P^* \rangle }{\partial x} + \mu \left(\dfrac{\partial^2 \langle u \rangle }{\partial y^2} \right) 
\end{equation}

\noindent where $-\partial \langle P^* \rangle/ \partial x$ is an equivalent pressure gradient term defined as 

\begin{equation}
\begin{split}
    - \dfrac{\partial \langle P^* \rangle }{\partial x} = &- \dfrac{\partial \langle p \rangle }{\partial x} + \mu \dfrac{\partial^2 \langle u \rangle}{\partial x^2} + {\mathcal{Z}} + \rho \left( \langle u \rangle \dfrac{ \partial \langle u \rangle}{\partial x} +   \langle v  \rangle \dfrac{\partial \langle u  \rangle}{\partial y} - \dfrac{\partial \langle uu \rangle}{\partial x} - \dfrac{\partial \langle uv  \rangle}{\partial y}\right) .
\end{split}
\end{equation}

\section{Spanwise-averaged Navier-Stokes equations in the curved leading edge}\label{appC}

\begin{figure}
    \centering
    \includegraphics[width=1\linewidth]{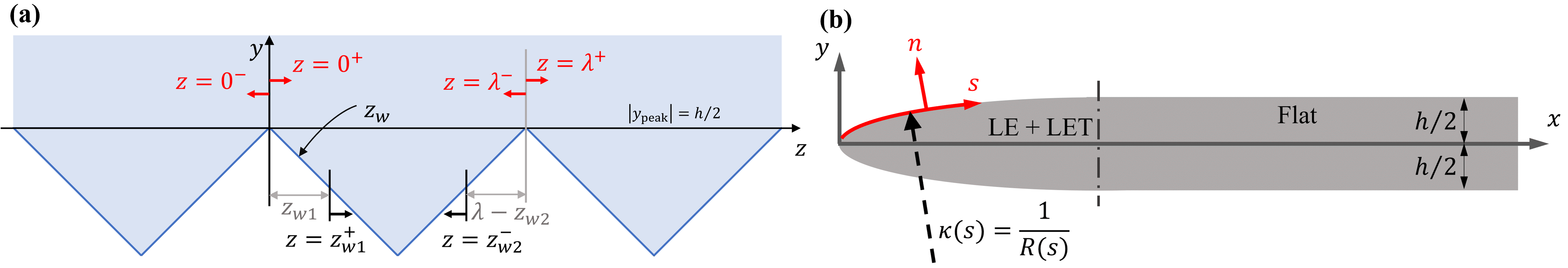}
    \caption{(a) Schematic of a riblet surface and the boundary conditions for characterizing the viscous diffusion in the $z$ direction as part of the averaging operation. (b) Schematic of the leading edge of the plate and the coordinate system tangent and normal to the wall. $\kappa(s)$ is the local curvature of the curved leading edge and is a function of the $s$ direction. }
    \label{fig:averaging_BC}
\end{figure}

For a surface with the local contour curvature defined as  $\kappa(s) = 1/R(s)$ and for an incompressible fluid \citep{schlichting2016boundary}, the Navier-Stokes equation in the direction tangent to the wall is written as (see figure \ref{fig:averaging_BC}(b)) 

\begin{equation}
    \begin{split}
    &\rho \left( \dfrac{1}{1+ \kappa n} u_s \dfrac{\partial u_s}{\partial s} + v_n \dfrac{\partial u_s}{\partial n}  + \dfrac{\kappa}{1+\kappa n} u_s v_n + w \dfrac{\partial u_s}{\partial z}    \right)  = \\ & -\dfrac{1}{1+ \kappa n} \dfrac{\partial p}{\partial s}  + 
     \dfrac{1}{(1 + \kappa n)}\left( \dfrac{\partial \tau_{ss}}{\partial s} +  \dfrac{1}{(1 + \kappa n)} \dfrac{\partial }{\partial n} \left[(1 + \kappa n)^2\tau_{sn}\right]\right) + \mu \dfrac{\partial^2 u_s}{\partial z^2} \label{NSE_curve}
    \end{split}
\end{equation}

\noindent where

\begin{equation}
    \tau_{ss} = \dfrac{2\mu}{1+\kappa n} \left( \dfrac{\partial u_s}{\partial s} + \kappa v_n \right)
    \label{t_ss}
\end{equation}

\noindent and

\begin{equation}
    \tau_{sn} = \mu \left( \dfrac{\partial u_s}{\partial n} - \dfrac{\kappa u_s}{1 + \kappa n} + \dfrac{1}{1 + \kappa n} \dfrac{\partial v_n}{\partial s} \right)
    \label{t_sn}
\end{equation}

\noindent First, in the absence of riblets, $w = 0$, ${\partial u_s}/{\partial z}= 0$, and ${\partial^2 u_s}/{\partial z^2}= 0$, and the equations return back to 2D forms. Substituting equations \eqref{t_ss} and \eqref{t_sn} in the viscous diffusion terms

\begin{equation}
    \dfrac{\partial \tau_{ss}}{\partial s} = \dfrac{2\mu}{(1+ \kappa n)} \dfrac{\partial^2 u_s}{\partial s^2} + \dfrac{2 \mu \kappa }{(1+ \kappa n)} \dfrac{\partial v_n}{\partial s} + 2 \mu {\mathcal{K}'} \label{viscou1}
\end{equation}

\noindent where ${\mathcal{K}'}$ combines all the terms that include the effect of the $\partial \kappa/\partial s$:

\begin{equation}
    {\mathcal{K}'} =  \dfrac{1}{(1+\kappa n)^2} \left(- n \dfrac{\partial u_s}{\partial s} + v_n\right) \dfrac{\partial \kappa}{\partial s}
\end{equation}

\noindent and 

\begin{equation}
    \begin{split}
    & \left( \dfrac{1}{(1 + \kappa n)} \dfrac{\partial }{\partial n} [(1 + \kappa n)^2\tau_{sn}] \right)  =  (1 + \kappa n) \dfrac{\partial \tau_{sn}}{\partial n} + 2\kappa \tau_{sn}  = \\
    & \mu (1 + \kappa n) \dfrac{\partial^2 u_s}{\partial n^2} + \mu \kappa \dfrac{\partial u_s}{\partial n} - \dfrac{\kappa^2 \mu}{1 + \kappa n} u_s + \dfrac{\mu \kappa}{(1+\kappa n)} \dfrac{\partial v_n}{\partial s} + \mu\dfrac{\partial^2 v_s}{\partial s \partial n} \label{viscous2}
    % \dfrac{\mu \kappa^2}{(1+\kappa n)} u_s + \mu\dfrac{\partial^2 v_s}{\partial s \partial n} - \dfrac{\mu \kappa}{(1+\kappa n)} \dfrac{\partial v_n}{\partial s} + \\
    % & 2 \kappa \mu \dfrac{\partial u_s}{\partial n} - \dfrac{2 \kappa^2 \mu}{1 + \kappa n} u_s + \dfrac{2 \kappa \mu}{1 + \kappa n} \dfrac{\partial v_n}{\partial s}
    \end{split}
\end{equation}

\noindent Adding the two viscous terms of equations \eqref{viscou1} and \eqref{viscous2} together and collecting and reorganizing some of the terms and dividing by $(1 + \kappa n)$ 

\begin{equation}
\begin{split}
    &\mu \left( \dfrac{\partial^2 u_s}{\partial s^2} + \dfrac{\partial^2 u_s}{\partial n^2} \right) +  
    \dfrac{\mu}{(1+\kappa n)} \dfrac{\partial}{\partial s} \left(\dfrac{1}{1+\kappa n} \dfrac{\partial u_s}{\partial s} + \dfrac{\partial v_n}{\partial n} + \dfrac{\kappa}{1+\kappa n} v_n \right) +  \dfrac{\mu {\mathcal{K}'}}{(1+\kappa n)} + \\ 
    &\mu \left( \dfrac{(-2\kappa n - \kappa^2 n^2)}{(1 + \kappa n)^2}\dfrac{\partial^2 u_s}{\partial s^2} + \dfrac{\kappa}{(1+\kappa n)} \dfrac{\partial u_s}{\partial n} +\dfrac{2 \kappa}{(1+\kappa n)^2} \dfrac{\partial v_n}{\partial s} - \dfrac{ \kappa^2}{(1 + \kappa n)^2} u_s
    \right)
    % + \dfrac{\kappa \mu}{1 + \kappa n} \dfrac{\partial v_n}{\partial s} + \\ & \ - \mu \kappa \dfrac{\partial u_s}{\partial n} + \dfrac{\mu \kappa^2}{(1+\kappa n)} u_s  - \dfrac{\mu \kappa}{(1+\kappa n)} \dfrac{\partial v_n}{\partial s} + \\ & 2 \kappa \mu \dfrac{\partial u_s}{\partial n} - \dfrac{2 \kappa^2 \mu}{1 + \kappa n} u_s + \dfrac{2 \kappa \mu}{1 + \kappa n} \dfrac{\partial v_n}{\partial s}
\end{split} \label{average_RHS_curved}
\end{equation}

\noindent where continuity dictates that 

\begin{equation}
    \left(\dfrac{1}{1+\kappa n} \dfrac{\partial u_s}{\partial s} + \dfrac{\partial v_n}{\partial n} + \dfrac{\kappa}{1+\kappa n} v_n \right) = -\dfrac{\partial w}{\partial z} \label{continuity_curve}
\end{equation}

\noindent and for 2D flows, $\partial w/\partial z = 0$. As for the pressure gradient term:

\begin{equation}
    -\dfrac{1}{1+ \kappa n} \dfrac{\partial p}{\partial s} = - \dfrac{\partial p}{\partial s} + \dfrac{\kappa n}{1+ \kappa n} \dfrac{\partial p}{\partial s}
\end{equation}

\noindent and thus the right hand side is simplified to

\begin{equation}
    \begin{split}
        &- \dfrac{\partial p}{\partial s} + \mu \left( \dfrac{\partial^2 u_s}{\partial s^2} + \dfrac{\partial^2 u_s}{\partial n^2} \right) + \dfrac{\kappa n}{1+ \kappa n} \dfrac{\partial p}{\partial s} +  \dfrac{\mu {\mathcal{K}'}}{(1+\kappa n)} + \\ 
        &\mu \left( \dfrac{(-2\kappa n - \kappa^2 n^2)}{(1 + \kappa n)^2}\dfrac{\partial^2 u_s}{\partial s^2} + \dfrac{\kappa}{(1+\kappa n)} \dfrac{\partial u_s}{\partial n} +\dfrac{2 \kappa}{(1+\kappa n)^2} \dfrac{\partial v_n}{\partial s} - \dfrac{ \kappa^2}{(1 + \kappa n)^2} u_s
    \right) .
    \end{split}
\end{equation}

\noindent As for the left hand side, we can rewrite the terms in the form:

\begin{equation}
    \begin{split}
        &\rho \left( \dfrac{1}{1+ \kappa n} u_s \dfrac{\partial u_s}{\partial s} + v_n \dfrac{\partial u_s}{\partial n}  + \dfrac{\kappa}{1+\kappa n} u_s v_n    \right) = \\ &\rho \left(u_s \dfrac{\partial u_s}{\partial s} + v_n \dfrac{\partial u_s}{\partial n} \right) - \rho \left( \dfrac{\kappa n}{1 + \kappa n} u_s \dfrac{\partial u_s}{\partial s} - \dfrac{\kappa}{1 + \kappa n} u_s v_n \right) 
    \end{split}
\end{equation}

\noindent Thus equation \eqref{NSE_curve} is re-arranged as 

\begin{equation}
    \rho \left(u_s \dfrac{\partial u_s}{\partial s} + v_n \dfrac{\partial u_s}{\partial n} \right) = - \dfrac{\partial p}{\partial s} + \mu \left(\dfrac{\partial^2 u_s}{\partial s^2} + \dfrac{\partial^2 u_s}{\partial n^2} \right) + {\mathcal{K}_1}  \label{BL_curvature_all}
\end{equation}

\noindent where ${\mathcal{K}_1}$ captures all the terms involving the curvature term $\kappa$:

\begin{equation}
\begin{split}
    {\mathcal{K}_1} = & \rho \left( \dfrac{\kappa n}{1 + \kappa n} u_s \dfrac{\partial u_s}{\partial s} - \dfrac{\kappa}{1 + \kappa n} u_s v_n \right) + \dfrac{\kappa n}{1+ \kappa n} \dfrac{\partial p}{\partial s} + \dfrac{\mu {\mathcal{K}'}}{(1+\kappa n)} + \\ & \mu \left( \dfrac{(-2\kappa n - \kappa^2 n^2)}{(1 + \kappa n)^2}\dfrac{\partial^2 u_s}{\partial s^2} + \dfrac{\kappa}{(1+\kappa n)} \dfrac{\partial u_s}{\partial n} +\dfrac{2 \kappa}{(1+\kappa n)^2} \dfrac{\partial v_n}{\partial s} - \dfrac{ \kappa^2}{(1 + \kappa n)^2} u_s \right).
    \label{BL_curvature}
\end{split}
\end{equation}

\noindent Similar to appendix \ref{appB} we can rewrite equation \eqref{BL_curvature_all} in the form of 

\begin{equation}
    \rho \left(u_s \dfrac{\partial u_s}{\partial s} + v_n \dfrac{\partial u_s}{\partial n} \right) = - \dfrac{\partial P^*}{\partial s} + \mu \left(\dfrac{\partial^2 u_s}{\partial n^2} \right) 
\end{equation}

\noindent where the equivalent pressure gradient term is 

\begin{equation}
    - \dfrac{\partial P^*}{\partial s} = - \dfrac{\partial p}{\partial s} + \mu \left(\dfrac{\partial^2 u_s}{\partial s^2} \right) + {\mathcal{K}_1} 
\end{equation}

\noindent and in the absence of curvature, $\kappa \rightarrow 0$ and $\partial \kappa / \partial s \rightarrow 0$, ${\mathcal{K}_1}$ becomes zero and the equations return back to that of the BL over the flat surface. Hence, in the curved region of the elliptical leading edge,  prior to the appearance of the riblets, the $-{\partial P^*}/{\partial s}$ term includes contributions from the pressure gradient, the viscous diffusion terms in the streamwise direction, as well as the curvature related terms as shown above and thus the $m$ parameter of the FS fit for the velocity profiles will capture the effect of these components.

Now for the textured portion of the curved leading edge, or LET, we apply the spanwise-averaging operation to the right hand side of equation \eqref{NSE_curve}, and cannot neglect the effect of the out of plane components:

\begin{equation}
\begin{split}
    & \left \langle \rho \left( \dfrac{1}{1+ \kappa n} u_s \dfrac{\partial u_s}{\partial s} + v_n \dfrac{\partial u_s}{\partial n}  + \dfrac{\kappa}{1+\kappa n} u_s v_n + w \dfrac{\partial u_s}{\partial z}    \right) \right\rangle = \\ &\rho \left( \dfrac{1}{1 + \kappa n} \left\langle u_s \dfrac{\partial u_s}{\partial s} \right\rangle + \left\langle v_n \dfrac{\partial u_s}{\partial n} \right\rangle + \left\langle w \dfrac{\partial u_s}{\partial z} \right\rangle + \dfrac{\kappa}{1+\kappa n} \langle u_s v_n \rangle \right)  \label{RHS_average}
\end{split}
\end{equation}

\noindent where using the product rule we substitute the first two terms with 

\begin{equation}
    \left \langle u_s \dfrac{\partial u_s}{\partial s} \right \rangle = \dfrac{\partial \langle u_s u_s \rangle }{\partial s} - \left \langle u_s \dfrac{\partial u_s}{\partial s} \right \rangle, \label{chainrule1}
\end{equation}

\begin{equation}
    \left \langle v_n \dfrac{\partial u_s}{\partial n} \right \rangle = \dfrac{\langle \partial u_s v_n \rangle}{\partial s} - \left \langle u_s \dfrac{\partial v_n}{\partial n} \right \rangle \label{chainrule2}
\end{equation}

\begin{equation}
    \left \langle w \dfrac{\partial u_s}{\partial z} \right \rangle = \left \langle \dfrac{\partial u_s w}{\partial s} \right \rangle - \left \langle u_s \dfrac{\partial w}{\partial z} \right \rangle \label{chainrule3}
\end{equation}

\noindent where using similar integral operation as discussed in appendix \ref{appB} in equation \eqref{uw_0}, 

\begin{equation}
    \left \langle \dfrac{\partial u_s w}{\partial s} \right \rangle = 0
\end{equation}

\noindent Then, multiplying the continuity equation with $u_s$ and applying the spanwise-averaging operation we can write:

\begin{equation}
    \dfrac{1}{1+\kappa n} \left\langle u_s \dfrac{\partial u_s}{\partial s} \right\rangle + \left\langle u_s \dfrac{\partial v_n}{\partial n} \right\rangle + \left\langle u_s \dfrac{\partial w}{\partial z} \right\rangle = - \dfrac{\kappa}{1 + \kappa n} \langle u_s v_n \rangle \label{continuty_average}
\end{equation}  
   
\noindent and thus using equations \eqref{chainrule1}, \eqref{chainrule2}, and \eqref{continuty_average}, we rewrite  \eqref{RHS_average} as 

\begin{equation}
    \rho \left( \dfrac{1}{1+ \kappa n}  \dfrac{\partial \langle u_s u_s \rangle}{\partial s} + \dfrac{\partial \langle u_s v_n \rangle}{\partial n} + \dfrac{2 \kappa}{1 + \kappa n} \langle u_s v_n \rangle \right)
\end{equation}

\noindent which similar to earlier it can be divided into:

\begin{equation}
    \rho \left( \dfrac{\partial \langle u_s u_s \rangle}{\partial s} + \dfrac{\partial \langle u_s v_n \rangle}{\partial n}  \right) + \rho \left(-\dfrac{\kappa n}{1 + \kappa n}   \dfrac{\partial \langle u_s u_s \rangle}{\partial s} + \dfrac{2 \kappa}{1 + \kappa n} \langle u_s v_n \rangle\right)
\end{equation}

\noindent As for the left hand side, following similar steps as before, using equations \eqref{average_RHS_curved} and \eqref{continuity_curve}, we can write the viscous terms as

\begin{equation}
    \begin{split}
    & \left \langle \dfrac{1}{(1 + \kappa n)}\left( \dfrac{\partial \tau_{ss}}{\partial s} +  \dfrac{1}{(1 + \kappa n)} \dfrac{\partial }{\partial n} [(1 + \kappa n)^2\tau_{sn}] \right)+ \mu \dfrac{\partial^2 u_s}{\partial z^2} \right \rangle = \\ & \mu \left( \dfrac{\partial^2 \langle u_s \rangle}{\partial s^2}  + \dfrac{\partial^2 \langle u_s \rangle}{\partial n^2} + \left \langle \dfrac{\partial^2 u_s}{\partial z^2} \right \rangle \right) - \dfrac{\mu}{1+\kappa n} \dfrac{\partial}{\partial s} \left\langle \dfrac{\partial w}{\partial z} \right\rangle +  \dfrac{\mu \langle {\mathcal{K}'} \rangle}{(1+\kappa n)} +  \\ & \mu \left( \dfrac{(-2\kappa n - \kappa^2 n^2)}{(1 + \kappa n)^2}\dfrac{\partial^2 \langle u_s \rangle}{\partial s^2} + \dfrac{\kappa}{(1+\kappa n)} \dfrac{\partial \langle u_s \rangle}{\partial n} +\dfrac{2 \kappa}{(1+\kappa n)^2} \dfrac{\partial \langle v_n \rangle}{\partial s} - \dfrac{ \kappa^2}{(1 + \kappa n)^2} \langle u_s \rangle
        \right)
    \end{split}
\end{equation}

\noindent where

\begin{equation}
    \left\langle \dfrac{\partial w}{\partial z} \right\rangle =\dfrac{1}{\lambda}\int_0^{\lambda} \dfrac{\partial w}{\partial z} dz = w(\lambda) - w(0) = 0
\end{equation}

\noindent and similar to \eqref{viscousBCZ} in appendix \ref{appB}

\begin{equation}
    \mu \left\langle \dfrac{\partial^2 u_s}{\partial z^2} \right\rangle = {\mathcal{Z}} = \left\{
    \begin{alignedat}{3} 
        & - \dfrac{2\mu}{\lambda} \dfrac{\partial u_s}{\partial z} \bigg \vert_{z=z_{\rm w1}^+}  \ \ \ \ \ & n_{\rm trough} \leq n \leq n_{\rm peak} \\
        &0 \ \ \ \ \ &\vert n \vert  > n_{\rm peak}.
    \end{alignedat}
    \right. 
\end{equation}

\noindent As for the pressure gradient term similarly:

\begin{equation}
    -\dfrac{1}{1+ \kappa n} \dfrac{\partial \langle p \rangle
    }{\partial s} = - \dfrac{\partial \langle p \rangle}{\partial s} + \dfrac{\kappa n}{1+ \kappa n} \dfrac{\partial \langle p \rangle}{\partial s}
\end{equation}

\noindent and thus 

\begin{equation}
    \begin{split}
    \rho \left( \langle u_s \rangle \dfrac{\partial \langle u_s \rangle}{\partial s} + \langle v_n \rangle \dfrac{\partial \langle u_s \rangle}{\partial n}  \right)
    =  & - \dfrac{\partial \langle P^* \rangle}{\partial s} + \mu \left( \dfrac{\partial^2 \langle u_s \rangle}{\partial n^2} \right) 
    \end{split}
\end{equation}

\noindent where 

\begin{equation}
    \begin{split}
    - \dfrac{\langle P^* \rangle}{\partial s} =  & - \dfrac{\partial \langle p \rangle}{\partial s}  + \mu \left( \dfrac{\partial^2 \langle u_s \rangle}{\partial s^2} \right) + {\mathcal{Z}} +  {\mathcal{K}_2} +  \\ &\rho \left( \langle u_s \rangle \dfrac{\partial \langle u_s \rangle}{\partial s} + \langle v_n \rangle \dfrac{\partial \langle u_s \rangle}{\partial n} - \dfrac{\partial \langle u_s u_s \rangle}{\partial s} -  \dfrac{\partial \langle u_s v_n \rangle}{\partial n}  \right) 
    \end{split}
\end{equation} 

\noindent and 

\begin{equation}
\begin{split}
    {\mathcal{K}_2}  = & \rho \left(\dfrac{\kappa n}{1 + \kappa n}   \dfrac{\partial \langle u_s u_s \rangle}{\partial s} - \dfrac{2 \kappa}{1 + \kappa n} \langle u_s v_n \rangle\right) + \dfrac{\kappa n}{1+ \kappa n} \dfrac{\partial \langle p \rangle}{\partial s} + \dfrac{\mu \langle {\mathcal{K}'} \rangle}{(1+\kappa n)} +\\ & \mu \left( \dfrac{(-2\kappa n - \kappa^2 n^2)}{(1 + \kappa n)^2}\dfrac{\partial^2 \langle u_s \rangle}{\partial s^2} + \dfrac{\kappa}{(1+\kappa n)} \dfrac{\partial \langle u_s \rangle}{\partial n} +\dfrac{2 \kappa}{(1+\kappa n)^2} \dfrac{\partial \langle v_n \rangle}{\partial s} - \dfrac{ \kappa^2}{(1 + \kappa n)^2} \langle u_s \rangle  \right) = \\ & \langle {\mathcal{K}_1} \rangle - \dfrac{ \rho \kappa}{1 + \kappa n} \langle u_s v_n \rangle
       .
\end{split}
\end{equation}

\bibliographystyle{jfm}
\bibliography{jfm-instructions}

\end{document}